\documentclass[aps,prl,twocolumn,nofootinbib,superscriptaddress]{revtex4-2}
\usepackage[dvipdfmx]{graphicx}
\usepackage{bm}
\usepackage{amsmath}
\usepackage{amsfonts}
\usepackage{mathrsfs}
\usepackage{listings}
\usepackage{color}
\usepackage{braket}
\pdfoutput=1

\newcommand{\beq}{\begin{eqnarray}}
\newcommand{\eeq}{\end{eqnarray}}

\usepackage[normalem]{ulem}
\newcommand\erase{\bgroup\markoverwith{\textcolor{blue}{\rule[.5ex]{2pt}{1pt}}}\ULon}

\usepackage{hyperref}

\begin{document}


\title{Spin-Nernst Effect in Time-Reversal-Invariant Topological Superconductors}


\author{Taiki Matsushita}
\affiliation{Department of Materials Engineering Science, Osaka University, Toyonaka, Osaka 560-8531, Japan}
\author{Jiei Ando}
\affiliation{Department of Materials Engineering Science, Osaka University, Toyonaka, Osaka 560-8531, Japan}
\author{Yusuke Masaki}
\affiliation{Department of Applied Physics, Graduate School of Engineering, Tohoku University, Sendai, Miyagi 980-8579, Japan}
\author{Takeshi Mizushima}
\affiliation{Department of Materials Engineering Science, Osaka University, Toyonaka, Osaka 560-8531, Japan}
\author{Satoshi Fujimoto}
\affiliation{Department of Materials Engineering Science, Osaka University, Toyonaka, Osaka 560-8531, Japan}
\affiliation{Center for Quantum Information and Quantum Biology, Osaka University, Toyonaka, Osaka 560-8531, Japan}
\author{Ilya Vekhter}
\affiliation{Department of Physics and Astronomy, Louisiana State University, Baton Rouge, LA 70803-4001, USA}

\date{\today}

\begin{abstract}
We investigate the spin-Nernst effect in time-reversal invariant topological superconductors, and show that it provides a smoking-gun evidence for helical Cooper pairs.
The spin-Nernst effect stems from asymmetric, in spin space, scattering of quasiparticles at nonmagnetic impurities, and generates a transverse spin current by the temperature gradient.
Both the sign and the magnitude of the effect sensitively depend on the scattering phase shift at impurity sites.
Therefore the spin-Nernst effect is uniquely suitable for identifying time-reversal invariant topological superconducting orders.
\end{abstract}
\maketitle

{\it Introduction.---}
In the last decade many researchers investigated topological superconductors (TSCs) with an eye on their application to future technologies such as quantum computation and spintronics~\cite{kitaev2001unpaired,qi2011topological,fu2008superconducting,linder2015superconducting,wakamura_spinpump,PhysRevLett.112.036602,PhysRevB.100.024502}.
These materials are characterized by non-trivial topology of the quasiparticle wave functions, which are responsible for the existence of Majorana quasiparticles~\cite{alicea2012new,sato_majorana,sato_topological}.
The signatures of non-trivial topology appear in transport phenomena, including the quantization of thermal Hall conductivity and tunneling conductance~\cite{read2000paired,nomura2012cross,sumiyoshi_ATHE,KTlow}.

TSCs can be divided into subclasses according to their behavior under time-reversal.
Condensate in TSCs with spontaneously broken time-reversal symmetry is formed from chiral Cooper pairs with a fixed angular momentum.
The key ingredients of time-reversal invariant (TRI) TSCs are helical Cooper pairs, which are equal mixtures of time-reversed copies of chiral Cooper pairs.
If TRI TSCs have an additional discrete symmetry, such as mirror plane, this symmetry may protect a pair of nodal points in the superconducting gap.
In the vicinity of each node, the low-energy Bogoliubov quasiparticles behave as Dirac fermions, and the corresponding class of materials is referred to as Dirac superconductors (DSCs).
The B-phase of the superfluid $^3$He, the fully gapped Balian--Werthamer (BW) state, where all three spin components of the triplet order parameter occur in equal measure, is a prototype of TRI TSCs~\cite{leggett1975theoretical,mizushima_He,volovic}.
In superconducting materials, there are several candidates for TRI TSCs and DSCs, including
$M_x$Bi$_2$Se$_3$ ($M={\rm Cu}, {\rm Sr}, {\rm Nb}$)~\cite{wray2010observation,kriener2011electrochemical,kriener2011bulk,matano2016spin,fu_CUBi2Se3,hor2010superconductivity,sasaki2011topological,yonezawa2017thermodynamic,sasakiPC15,yonezawa2016bulk,yonezawa,uematsu2019chiral,pan2016rotational,nikitin2016high}, U$_{1-x}$Th$_x$Be$_{13}$~\cite{machida2018spin,mizushima2018topology,shimizu2017quasiparticle,sigrist_UBe13,Heffner_UBe13}, and Cd$_3$As$_2$~\cite{hashimoto2016superconductivity,kobayashi2015topological,matsushita2018charge}.
Even though full range of experimental probes has been used for these compounds, including heat capacity, thermal transport, nuclear magnetic resonance, tunneling spectroscopy and other measurements, the unequivocal ``smoking gun'' evidence for TSCs/DSCs remains elusive.
Hence, it is indispensable to elucidate physical properties directly associated with helical Cooper pairs.

\begin{figure}[t]
\includegraphics[width=85mm]{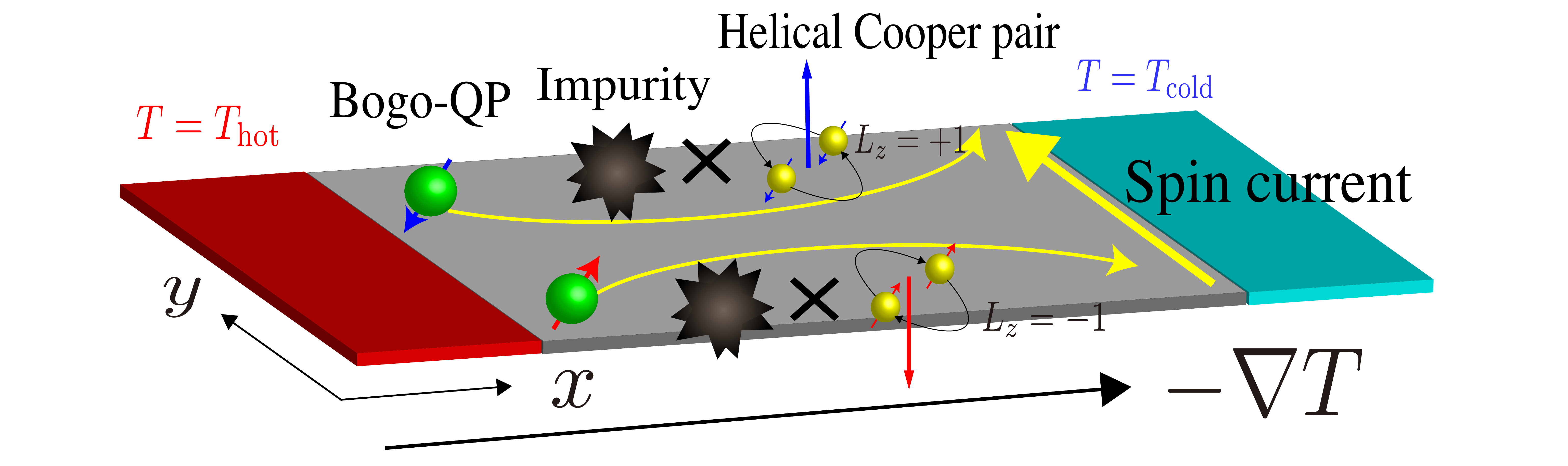}
\caption{Schematic image of the SNE in TRI TSCs : spin-dependent asymmetric scattering of the quasiparticles reflecting coupling to different angular momentum components of the Cooper pairs leads to transverse spin current.
}
\label{fig1}
\end{figure}

Among various transport phenomena, the Nernst effect, the transverse electric field generated by a thermal gradient in the presence of a magnetic field, is a powerful tool to capture the symmetry of superconducting order parameters~\cite{behnia2016nernst}.
The Nernst effect induced by flux flow and superconducting fluctuations has been extensively investigated in a variety of materials~\cite{zeh1990nernst,wang2006nernst,ussishkin2002gaussian,ussishkin2003superconducting,kontani2002nernst,pourret2006observation,zhang2008anomalous}.
In URu$_2$Si$_2$, the giant Nernst effect observed above the superconducting transition temperature was attributed to the fluctuations of preformed chiral Cooper pairs~\cite{sumiyoshi2014giant,yamashita2015colossal}.
Thus, the Nernst effect provides a direct probe for Cooper pairs with spontaneously broken time-reversal symmetry.

In this Letter, we show that the {\em spin Nernst effect} (SNE), the transverse spin current induced by a thermal gradient {\em{without}} an applied magnetic field, is a signature of TRI TSCs.
On the basis of the quasiclassical transport theory, we demonstrate that in TRI TSCs the impurity scattering of quasiparticle induces the SNE, which reflects the helical nature of the Cooper pairs, schematically shown in Fig.~\ref{fig1}.

We reiterate that the SNE in TRI TSCs essentially differs from the conventional Nernst effect due to superconducting fluctuations or vortex motion because: (i) A magnetic field is unnecessary, and (ii) the SNE is the bulk transport of homogeneous superconductors below the superconducting transition temperature, $T_c$.
As the SNE arises purely due to the symmetry of Cooper pairs, it provides smoking-gun evidence for helical Cooper pairs in TRI TSCs.


{\it Quasiclassical Keldysh transport theory.---}
The quasiclassical approximation, valid when $(k_{\rm F}\xi)^{-1}\ll 1$, where $\bm k_{\rm F}$ is the Fermi momentum and $\xi$ is the coherence length, is applicable to many superconductors and provides a powerful tool to investigate their transport properties.
These are determined from the quasiclassical Green's function $\check{g}(\epsilon,\bm x, \bm k_{\rm F})$, which is an $8\times 8$ matrix in the Keldysh and Nambu (particle-hole) space, defined for each {$\bm k_{\rm F}$ ~\cite{eilenberger1968transformation,serene1983quasiclassical}.
To leading order in $(k_{\rm F}\xi)^{-1}$} the Green's function obeys the quasiclassical Eilenberger equation~\cite{kobayashi2018negative},
\begin{eqnarray}
\label{Keldysh_eq}
&\left[ \epsilon\check{\tau}_z-\check{\Delta}-\check{\sigma}_{\rm imp},\check{g} \right]+i{\bm v}_{\rm F}\!\cdot\! \nabla \check{g}=0\,,
\end{eqnarray}
supplemented by the normalization condition
$\check{g}^2=-\pi^2$~\cite{SM,eilenberger1968transformation}.
In Eq.~\eqref{Keldysh_eq}, $\check{\tau}_i \; (i=x,y,z)$ are the Pauli matrices in the Nambu space, ${\bm v}_{\rm F}$ is the Fermi velocity of normal quasiparticles,
$\check{\sigma}_{\rm imp}$ is the impurity self-energy, and $\check{\Delta}$ is the superconducting order parameter matrix, determined from the self-consistency equation~\cite{sigrist1991phenomenological}.
We set $\hbar=k_{\rm B}=1$, and give the details of other notation and formulation in the supplemental material~\cite{SM}.

In the quasiclassical limit, the spin current is obtained from the Keldysh component, $\underline{g}^{\rm K}$, of the Green's function, $\check{g}$, as
\begin{align}
\label{spin_current}
{\bm J}^{\sigma_\mu}&=\frac{1}{2}N(\epsilon_{\rm F}) \int \frac{d\epsilon}{4\pi i}\left\langle{\rm Tr}\left[{\bm v_{\rm F}} \underline{\sigma}_\mu \underline{\tau}_z\underline{g}^{\rm K}\right]\right\rangle_{\rm FS},
\end{align}
where $N(\epsilon_{\rm F})$ is the normal-state density of states at the Fermi energy and $\underline{\sigma}_\mu \;(\mu=x,y,z)$ is the spin operator in the Nambu space.
$\langle\ldots\rangle_{\rm FS}$ denotes the normalized Fermi surface average.
We compute the linear, in the temperature gradient ($-{\bm \nabla}T$), correction to $\underline{g}^{\rm K}$, accounting for impurity scattering in the self-consistent $T$-matrix approximation (SCTA)~\cite{graf1996electronic,vorontsov2007unconventional}. When implemented in the Keldysh formalism, the anomalous self-energy contains the contribution of the vertex corrections~\cite{SM}.
These vertex corrections are essential for skew-scattering~\cite{nagaosa2010anomalous} and
generation of the transverse spin current defined in Eq.~\eqref{spin_current}.

In the following we assume the $\delta$-function individual impurity potential with the strength $V_{\rm imp}$, and the density of impurities $n_{\rm imp}$.
SCTA gives for the impurity self-energy
\begin{eqnarray}
\label{selfen}
\check{\sigma}_{\rm imp}(\epsilon)=-\Gamma_{\rm imp} \left(\cot \delta+\left\langle\frac{ \check{g}}{\pi}\right\rangle_{\rm FS} \right)^{-1}.
\end{eqnarray}
Here, we defined the normal-state scattering rate $\Gamma_{\rm imp}=\frac{n_{\rm imp}}{\pi N(\epsilon_{\rm F})}$ and the scattering phase shift
$\cot\delta = -1/[\pi N(\epsilon_{\rm F})V_{\rm imp}]$.
The limit $\delta\rightarrow 0$ ($\delta\rightarrow\pi/2$) corresponds to the Born (unitarity) scattering.
We then compute the tensor of spin-Nernst coefficients (SNCs), $\alpha_{jl}^{\sigma_\mu}$, from the linear response to the thermal gradient,
\begin{eqnarray}
J^{\sigma_\mu}_j=\alpha_{jl}^{\sigma_{\mu}}(-\partial_l T).
\end{eqnarray}
This expression neglects possible normal state spin-Nernst coefficient due to spin-orbit interaction, but this contribution is expected to be very small for $T_c\ll E_{SO}$, where $E_{SO}\sim 10^3$K is the characteristic spin-orbital energy scale.~\cite{Vorontsov:2008,Dyrdal:2016}

{\it SNE in DSCs.---}
As a prototype of TRI TSCs, we consider the three-dimensional helical $p$-wave superconducting gap on the spherical Fermi surface, where the $d$-vector is given by
\begin{align}
\label{dvec_DSC}
{\bm d}_{{\rm DSC},xy}(\bm k)=\frac{\Delta}{k_{\rm F}}(k_x,k_y,0).
\end{align}
This is an example of a DSC since the gap has two Dirac points (DPs) at the south and north poles on the Fermi sphere.
This simple gap structure enables one to capture essence of the SNE in TRI TSCs.
The same model describes the low-energy physics of DSCs in Cd$_3$As$_2$, and the results below are directly applicable to this material~\cite{matsushita2018charge}.
The spin projection ${\sigma_z}$ is a good quantum number for Eq.~(\ref{dvec_DSC}), and the quasiparticle states are block-diagonal in terms of $\sigma_z=\pm 1$.
The Cooper pairs in $\sigma_z=+1$ ($\sigma_z=-1$) sector condense into the $L_z=-1$ ($L_z=+1$) eigenstates of the angular momentum, $k_x-ik_y$ ($k_x+ik_y$).
Each sector is chiral and breaks time-reversal and mirror symmetries.
These broken symmetries give rise to the asymmetric quasiparticle scattering at impurities, which induces a transverse flow of quasiparticles along the direction determined by the sign of chirality~\cite{yip_ATHE,wave_ATHE,ikegami2013chiral,PhysRevB.94.064511}.
Since the helical pairing state or DSCs can be regarded as the superposition of chiral Cooper pairs with opposite chiralities in different spin sectors, asymmetric scattering on nonmagnetic impurities becomes spin-selective and thus generates the transverse spin current (Fig.~\ref{fig1}).

\begin{figure}[t]
\includegraphics[width=\columnwidth]{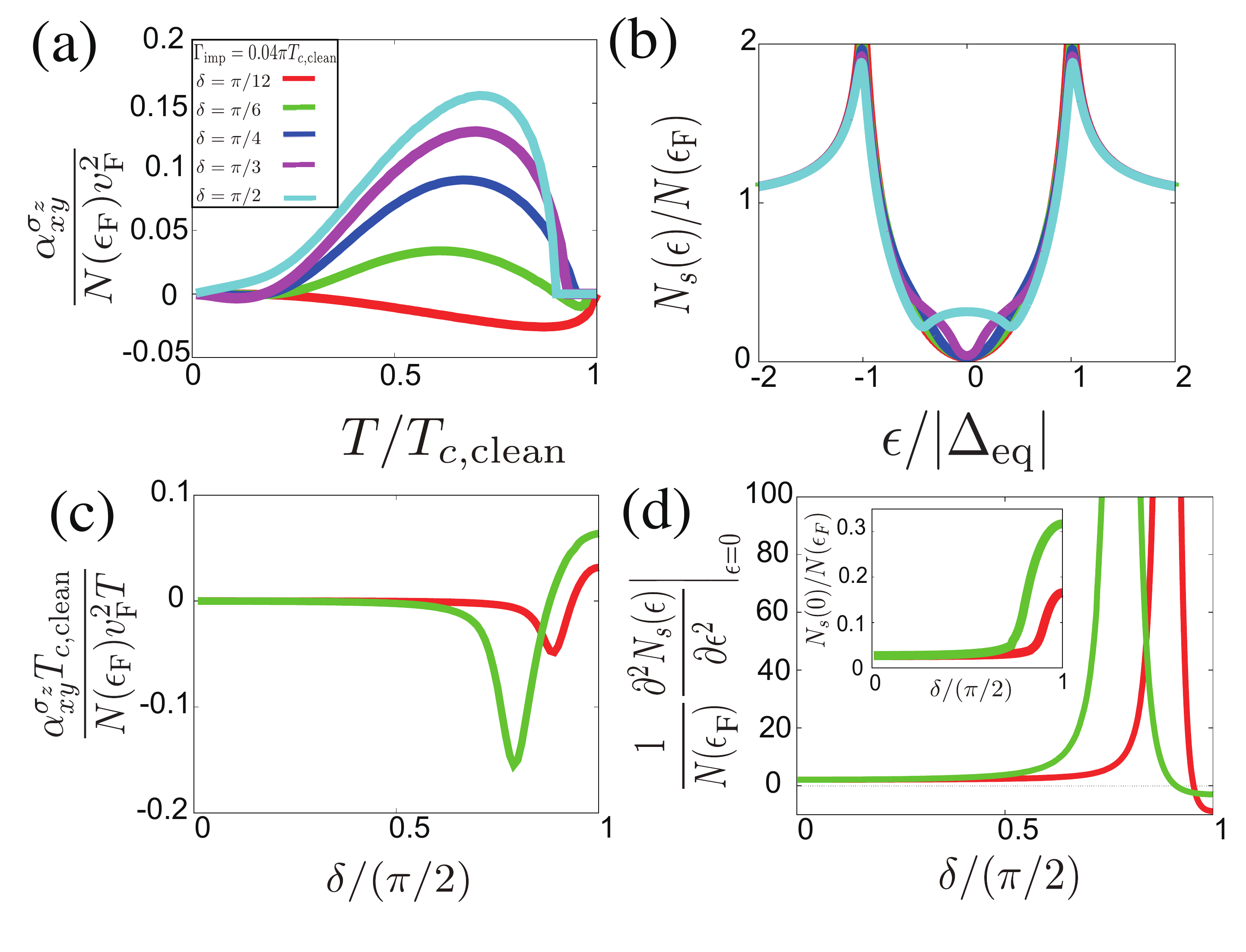}
\caption{
(a) Temperature dependences of the SNC ($\alpha_{xy}^{\sigma_z}$) and (b) quasiparticle DOS in DSCs at $T =0.01\pi T_{c,{\rm clean}}$ with a critical temperature at the clean limit, $T_{c,{\rm clean}}$.
For the panels~(a,b), we set the impurity scattering rate $\Gamma_{\rm imp} =0.04\pi T_{c,{\rm clean}}$, and the scattering phase shift $\delta=\frac{\pi}{12}$ (red), $\frac{\pi}{6}$ (green), $\frac{\pi}{4}$ (blue), $\frac{\pi}{3}$ (purple), $\frac{\pi}{2}$ (light blue).
(c) The scattering phase-shift dependence of $\alpha_{xy}^{\sigma_z}$ in DSCs at $T=0.01T_{c,{\rm clean}}$ and (d) the second-order derivative of the DOS with respect to $\epsilon$.
For the panels~(c,d), we set the impurity scattering rate $\Gamma_{\rm imp} =0.01\pi T_{c,{\rm clean}}$ for the red curves, and $\Gamma_{\rm imp} =0.04\pi T_{c,{\rm clean}}$ for the green curves, respectively.
The inset in the panel (d) shows the phase-shift dependence of the residual DOS $N_s(0)$.
}
\label{fig5}
\end{figure}

Motivated by this, we consider the spin-Nernst signal for the temperature gradient along the $y$-direction.
Fig.~\ref{fig5}(a) shows the temperature dependence of the spin-Nernst coefficient for different impurity scattering phase shifts.
The SNC is sensitive to the scattering phase shift.
Both the low-temperature slope and the maximum value below $T_c$ increase as the phase shift approaches the unitarity limit, $\delta \rightarrow \pi/2$.
Remarkably, the sign of the SNC changes as a function of $\delta$.

This evolution can be understood from the low-temperature expansion of $\underline{g}^{\rm K}$ in Eq.~\eqref{spin_current}~\cite{graf1996electronic}.
Since the nonequilibrium Keldysh Green's function is proportional to ${\rm sech}^2(\epsilon/2T)$, it entails a frequency cutoff $\epsilon \sim T$ that serves as a small parameter at low $T$.
We find for the SNC in clean DSCs as $T\rightarrow 0$~\cite{SM},
\begin{align}
\label{spinNernst_coef}
\frac{\alpha^{\sigma_z}_{xy}}{N(\epsilon_{\rm F})v_{\rm F}^2 }&=
-\frac{\pi^2\gamma \Gamma_{\rm imp} T }{12}
\frac{\cot^2 \delta-{n_s^2}}{\left(\cot^2 \delta +n_s^2\right)^2 }\nonumber\\
&\times \left\langle\frac{|{\bm d}_{\rm eq}({\bm k_{\rm F}})|^2}{[|{\bm d}_{\rm eq}({\bm k_{\rm F}})|^2+\gamma^2]^{3/2}}\right\rangle_{\rm FS}^2+
\mathcal{O}(T^2,\Gamma_{\rm imp}^3)\,.
\end{align}
The complete expression of the SNC, including higher-order terms for $\Gamma_{\rm imp}$, is given in the supplemental materials~\cite{SM}.
In Eq.~(\ref{spinNernst_coef}), we introduced the residual quasiparticle DOS at the Fermi energy in the superconducting state, $n_s=N_s(0)/N(\epsilon_{\rm F})=-(1/4\pi)\langle{\rm Tr\ Im}[\underline{\tau_z}\underline{g}^{\rm R}_{\rm eq}(\epsilon)]\rangle_{\rm FS}$, and the impurity self-energy at equilibrium, $\gamma \equiv \frac{i}{4}{\rm Tr}(\underline{\tau}_z\underline{\sigma}_{\rm imp,eq}^{\rm R}(0))$.
It is seen from Fig.~\ref{fig5}(b) that as the phase shift approaches the unitarity limit, the spectral weight around the coherence peaks, $\epsilon \approx \pm |\Delta_{\rm eq}|$ reduces, while $N_{ s}(0)$ increases.
The transfer of the spectral weight in the DOS reflects the formation of the impurity bands, see discussion below.

It is clear from Eq.~\eqref{spinNernst_coef} that, in agreement with Fig.~\ref{fig5}(a), the SNC changes the sign as a function of the scattering phase shift from negative at weak scattering, $\cot\delta\gg 1$ to positive near unitarity, $ \cot\delta\rightarrow 0$.
This behavior is shown in Fig.~\ref{fig5}(c), where the critical phase shift is $\delta_{\rm c} \simeq 0.94\times \frac{\pi}{2}\;$ for $\;\Gamma_{\rm imp}=0.01\pi T_{c,{\rm clean}}$, and $\delta_{\rm c} \simeq 0.88\times \frac{\pi}{2}\;$ for $\; \Gamma_{\rm imp}=0.04\pi T_{c,{\rm clean}}$.
Expansion in the phase shift at low energies near the unitarity limit gives the sign change occurring at $\delta_{\rm c}=(\pi/2)(1-\chi_{\rm c})$ with $\chi_{\rm c}\approx\sqrt{\Gamma_{\rm imp}/\pi \Delta_{\rm eq}(T=0)}$ in qualitative agreement with the numerical results.

Another striking feature in Fig.~\ref{fig5}(c) is a large peak in SNC at intermediate phase shift.
Recall that the transverse transport coefficients reflect the asymmetry of scattering convoluted with the variation of the density of states near the Fermi surface~\cite{Arfi1988}.
In unconventional superconductors, the impurity potential gives rise to the impurity quasi-bound (resonant) states, whose position shifts from the gap edge to mid-gap as the phase shifts approaches the unitarity limit [see Fig.~\ref{fig5}(b)].
For finite impurity concentrations, these states (symmetrically positioned at the electron- and hole-sides of the spectrum) broaden into the impurity bands.
Sizeable DOS at the Fermi level first appears when the impurity band touches the Fermi energy. At that point the strong variation of the
the DOS with the energy amplifies the scattering asymmetry near the Fermi energy. 
A quantitative measure of when this happens is the band curvature at origin, $\frac{d^2N_s}{d\epsilon^2}\big|_{\epsilon=0}$, which is maximal when the bands first reach $\epsilon=0$.
The corresponding phase-shift is estimated to be $\delta=(\pi/2)(1-2\chi_{\rm c})$.
As shown in Fig.~\ref{fig5}(c,d), the peak in the band curvature coincides with the peak in SNC in Fig.~\ref{fig5}(c).

Since the mechanism for the SNE described here relies on the structure of the emergent impurity-induced bands, the same picture should be applicable to fully gapped TRI TSCs, which are considered below.

\begin{figure}[t!]
\includegraphics[width=9cm]{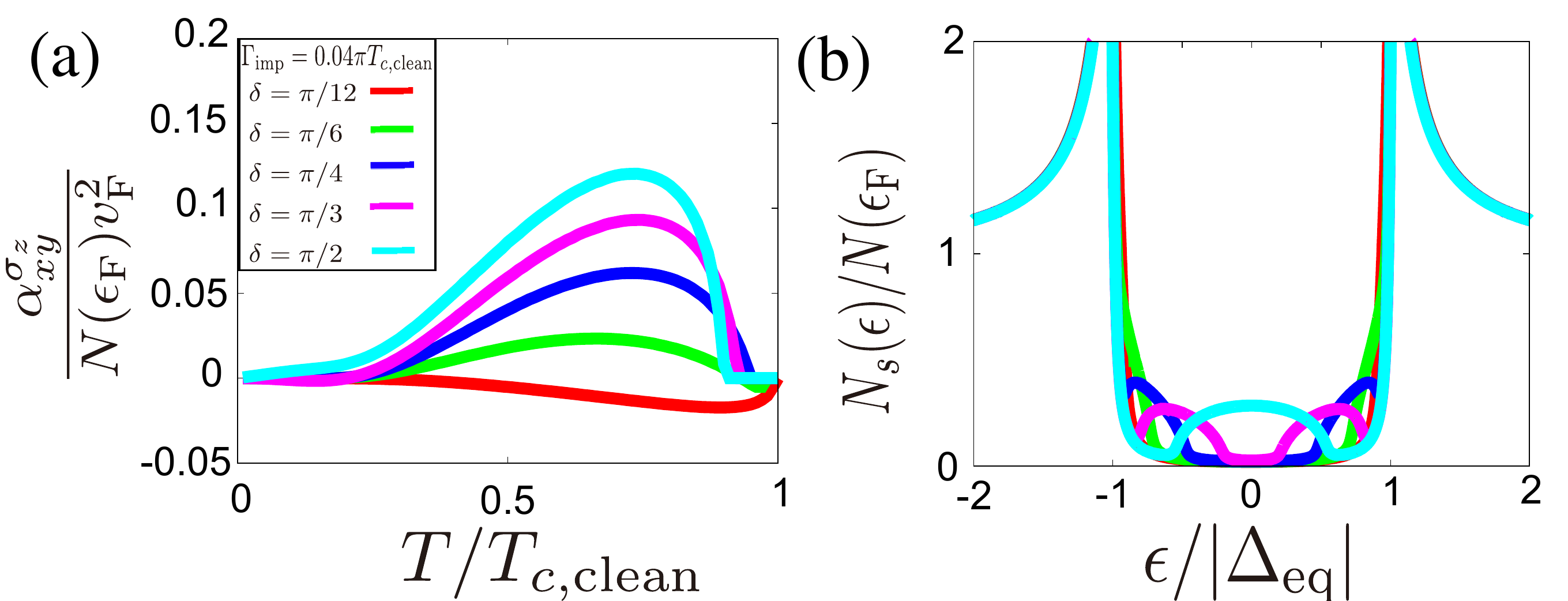}
\caption{
(a) The temperature dependences of the SNC and (b) the DOS in the dirty BW state at $T=0.01T_{c, {\rm clean}}$.
We set the impurity scattering rate $\Gamma_{\rm imp}=0.04\pi T_{c, {\rm clean}}$ and the phase shift $\delta=\frac{\pi}{12},\frac{\pi}{6},\frac{\pi}{4},\frac{\pi}{3},\frac{\pi}{2}$.
The unitarity limit $\delta \rightarrow \pi/2$ describes the gapless BW state.
}
\label{fig6}
\end{figure}

{\it The BW state in disordered media.---}
A well studied example of fully gapped TRI TSCs is the BW state, ${\bm d}_{{\rm BW}}(\bm k)=\frac{\Delta}{k_{\rm F}} (k_x,k_y,k_z)$, realized in the B-phase of the superfluid $^3$He~\cite{leggett1975theoretical,mizushima_He}.
Here we consider the BW state in the presence of nonmagnetic impurities.
At the qualitative level, the SNE in the BW state shares its origin with that in DSCs discussed above.
The BW state can be viewed as superposition of three helical $p$-wave pairing channels,
\begin{eqnarray}
\label{relation_BW_DSC}
{\bm d}_{\rm BW}=\frac{1}{2}\left({\bm d}_{{\rm DSC},xy}+{\bm d}_{{\rm DSC},yz}+{\bm d}_{{\rm DSC},zx}\right)\,,
\end{eqnarray}
with
${\bm d}_{{\rm DSC},xy}(\bm k)=\frac{\Delta}{k_{\rm F}}(k_x,k_y,0)$,
${\bm d}_{{\rm DSC},zx}(\bm k)=\frac{\Delta}{k_{\rm F}}(k_x,0,k_z)$,
and ${\bm d}_{{\rm DSC},yz}(\bm k)=\frac{\Delta}{k_{\rm F}}(0,k_y,k_z)$.
Figure~\ref{fig6} shows the temperature dependences of the SNC for several phase shifts, which are qualitatively same as those in DSCs discussed above.

The spin structure of these components gives rise to the SNE similar to the case of DSCs, with the result shown in
Fig.~\ref{fig6}(a).
The SO(3) symmetry preserved in the BW state dictates the relations between the tensor elements of the SNC,
\begin{eqnarray}
\alpha^{\sigma_z}_{xy}=\alpha^{\sigma_x}_{yz}=\alpha^{\sigma_y}_{zx}=-\alpha^{\sigma_z}_{yx}=-\alpha^{\sigma_x}_{zy}=-\alpha^{\sigma_y}_{xz}.
\label{eq:alpha}
\end{eqnarray}
These relations are also maintained by the $A_{u}$ irreducible representation of the $O_h$ crystals.
Eq.~\eqref{eq:alpha} can be understood from Eq.~(\ref{relation_BW_DSC}):
When the temperature gradient is applied along the $y$-direction, ${\bm d}_{{\rm DSC},xy}$ (${\bm d}_{{\rm DSC},yz}$) pairing gives rise to the spin current ${\bm J}^{\sigma_z}$ (${\bm J}^{\sigma_x}$) along the $x$-direction ($z$-direction).

As in DSCs, the quasiparticle DOS in the BW state with nonmagnetic impurities has impurity bound states, where the spectral weight is transferred from the coherence peaks around $\epsilon \approx \pm \Delta$ to the lower energies [Fig.~\ref{fig6}(b)].
The width of the band formed around these resonance energies depends on the phase shift of the scattering as well as the impurity concentration.
When the impurity bands reach $\epsilon=0$, the system realizes the ``gapless'' BW state.
Even though the BW state in the clean limit is fully gapped, the quasiparticles of the impurity bound states are responsible for the SNE.
Notably, in $\delta\rightarrow\pi/2$, the SNCs show the $T$-linear behavior at the low-temperature, which we attribute to the finite impurity-induced DOS at the Fermi level [see Figs.~\ref{fig6}(a,b)].

{\it Application to candidate materials.---}
A well-established example of gapped TRI TSCs is the B-phase of superfluid $^3$He.
For $^3$He, strong (near-unitarity) impurity scattering can be engineered by highly porous silica aerogel~\cite{halperin1,halperin2,halperin3}, realizing
the ``dirty'' BW state~\cite{thuneberg,priya1,priya2,higashitani2005microscopic}.
The aerogel is composed of silica strands (diameter $30 {\rm \AA}$), separated by the mean-distance that is comparable to or less than the superfluid coherence length.
The spin flip scattering by magnetic surface solid is suppressed by coating aerogel with a few layers of $^4$He atoms~\cite{sprague}.
Hence, the properties of the liquid $^3$He in the aerogel are well described by the homogeneous scattering model~\cite{thuneberg}, where the aerogel is represented by randomly distributed nonmagnetic scattering centers.
The model has two parameters: the phase shift $\delta$ and the mean free path $\ell$ determined by the aerogel geometry.
Several experiments identified the ``gapless'' BW state over the pressure range $p=6$-$34$ bar~\cite{choi,sauls05,fisher,yoon}, which is in good agreement with the homogeneous scattering model with $\delta \rightarrow \pi/2$ and $\ell\approx 1800\AA$ for $98\%$ porosity~\cite{thuneberg,priya1,priya2}.
Then the normal-state scattering rate for the aerogel is estimated as $\Gamma_{\rm imp}=\hbar v_F/(2 l \sin^2\delta)\approx 0.1$ - $0.2\pi T_{\rm c,clean}$.
We find the qualitatively same behavior of the SNC as that in Fig.~\ref{fig6}(a) even for such large $\Gamma_{\rm imp}$.
Hence, the spin-Nernst effect can be utilized as thermal generation of quasiparticle-mediated spin current in superfluid $^3$He.

Another interesting example is the heavy-fermion superconductor U$_{1-x}$Th$_x$Be$_{13}$, discovered in the 1980s~\cite{ott1983u}.
It is a spin-triplet superconductor with three distinct superconducting phases in the $x$-$T$ plane.
At $x=0$,the ``parent'' material UBe$_{13}$ undergoes superconducting phase transition at $T_{2c}(x=0) \sim 0.85$ K.
For $0\leq x< 0.02$, the critical temperature $T_{2c}(x)$ decreases with increasing Th concentration, $x$.
This superconducting phase is referred to as the C-phase.
In a narrow dopant region $0.02\leq x\leq 0.04$, an additional superconducting transition occurs at $T_{1c}(x) \geq T_{2c}(x)$, and the time-reversal symmetry is spontaneously broken below $T_{2c}(x)$~\cite{ott1985phase,kim1991investigation,PhysRevLett.55.1319}.
The superconducting phase in $T_{2c}(x)\leq T \leq T_{1c}(x)$ is referred to as the A-phase and the time-reversal symmetry broken phase is called the B-phase~\cite{shimizu2017quasiparticle,Heffner_UBe13}.

In spite of many efforts, the pairing symmetry of this material remains unresolved.
One possible scenario is an accidental degeneracy of the order parameters belonging to different irreducible representations of the $O_h$ group~\cite{sigrist_UBe13}.
Another possibility is the realization of the odd-parity $E_u$ state~\cite{mizushima2018topology}.
Both scenarios predict DSCs in the A-phase and TRI TSCs in the C-phase.

However, onset of the $E_u$ state is accompanied by a nematic phase transition with broken rotational symmetry, leading to a different type of helical Cooper pairing from the accidental scenario~\cite{machida2018spin}.
The Dirac superconducting A-phase supported by the accidental scenario manifests a finite SNC $\alpha_{xy}^{\sigma_z}$, $\alpha_{yz}^{\sigma_x}$ and $\alpha_{zx}^{\sigma_y}$, whereas only $\alpha_{xy}^{\sigma_z}$ is finite in the $E_u$ state.
For the TRI TSC C-phase, the nematicity in the $E_{u}$ state leads to the anisotropy of the SNC, while there is no anisotropy of the SNC in the same phase within the accidental scenario.
Thus, measurements of the SNE presented in this paper provide a smoking-gun evidence for identifying the superconducting symmetry in U$_{1-x}$Th$_x$Be$_{13}$ and other complex materials.

We note that in the weak coupling limit the form of the gap function fully determines the topological properties of the superconducting state (for a given Fermi surface). Therefore our results remain valid for systems where the parity of the superconducting state is determined, in real space, by orbital mixing, such as Cd$_3$As$_2$ and doped Bi$_2$Se$_3$. In the momentum space these order parameters map on the examples considered above.~\cite{yonezawa,hashimoto2016superconductivity}

{\it Conclusion.---}
We established that the SNE provides direct evidence for helical Cooper pairs in TRI TSCs.
The origin of this SNE is the spin-dependent scattering of quasiparticles through the helical Cooper pairs on nonmagnetic impurities.
The SNE has strong dependence on the scattering phase shift, and changes the sign of the SNC on approaching the unitarity limit.
The SNE is detectable in the TSC candidate materials and its experimental verification is feasible.

We finally comment on an interesting future study.
In this letter, we focused on the SNE in bulk but the SNE is also possible in the surfaces.
In the surfaces of the TRI TSCs, the low-energy quasiparticles behave as helical fermions and carry the circulating spin-current~\cite{TSC_qi}.
The SNE at the surfaces is expected through helical fermions or the parity mixing of the order parameters~\cite{sauls2011surface}.

\begin{acknowledgments}
T. Matsushita is grateful to A. Daido, A. Shitade and Y. Yanase for useful discussions.
Discussion with A. Daido for TRI TSC is one of the motivations for this research.
T. Matsushita was supported by a  Japan Society for the Promotion of Science (JSPS) Fellowship for Young Scientists and by JSPS KAKENHI Grant No.~JP19J20144 and Y.M was supported in part by the JSPS Early-Career Scientists Grant No.~JP19K14662. I.~V. was supported in part by NSF Grant DMR-1410741.
This work was initiated at Louisiana State University, and also supported by JST CREST Grant No. JPMJCR19T5, Japan, and the Grant-in-Aid for Scientific Research on Innovative Areas ``Quantum Liquid Crystals (JP20H05163)'' from JSPS of Japan, and JSPS KAKENHI (Grant No.~JP17K05517, No.~JP20K03860, No.~JP20H01857 and No. JP21H01039).
\end{acknowledgments}

\bibliography{thermospin.bib}
\bibliographystyle{apsrev4-1_PRX_style}
\end{document}



 \title[]
{
\large Supplemental Materials for \\
\smallskip
``Spin-Nernst Effect\\
in Time-Reversal-Invariant Topological Superconductors''
}
\author{Taiki Matsushita}
\affiliation{Department of Materials Engineering Science, Osaka University, Toyonaka, Osaka 560-8531, Japan}
\author{Jiei Ando}
\affiliation{Department of Materials Engineering Science, Osaka University, Toyonaka, Osaka 560-8531, Japan}
\author{Yusuke Masaki}
\affiliation{Department of Applied Physics, Graduate School of Engineering, Tohoku University, Sendai, Miyagi 980-8579, Japan}
\author{Takeshi Mizushima}
\affiliation{Department of Materials Engineering Science, Osaka University, Toyonaka, Osaka 560-8531, Japan}
\author{Satoshi Fujimoto}
\affiliation{Department of Materials Engineering Science, Osaka University, Toyonaka, Osaka 560-8531, Japan}
\affiliation{Center for Quantum Information and Quantum Biology, Osaka University, Toyonaka, Osaka 560-8531, Japan}
\author{Ilya Vekhter}
\affiliation{Department of Physics and Astronomy, Louisiana State University, Baton Rouge, LA 70803-4001}

\setcounter{equation}{0}
\setcounter{figure}{0}
\setcounter{table}{0}
\setcounter{page}{1}
\renewcommand{\theequation}{S\arabic{equation}}
\renewcommand{\thefigure}{S\arabic{figure}}
\renewcommand{\bibnumfmt}[1]{[S#1]}
\renewcommand{\citenumfont}[1]{S#1}


\maketitle
In this supplemental material, we present the quasiparticle transport theory, as well as the derivation of Eq.~(2), Eq.~(3), and Eq.~(6) in the main text.

\section{Quasiclassical transport equation for superconductors}
We begin by the Gor'kov equation in the Wigner representation~\cite{serene1983quasiclassical}.
The Green's function in this representation depends on the center of mass coordinate, $\bm x$, and the momentum corresponding to the relative motion, $\bm k$.
The Keldysh Green's function,
\begin{eqnarray}
\label{Greenfunc_Kspace}
\check{G}(\epsilon,\bm x,\bm k)
&=&
\begin{pmatrix}
\underline{G}^{{\rm R}}(\epsilon,\bm x,\bm k)&&\underline{G}^{{\rm K}}(\epsilon,\bm x,\bm k)\\
0&&\underline{G}^{{\rm A}}(\epsilon,\bm x,\bm k)
\end{pmatrix},\\
\label{Greenfunc_Nspace}
\underline{G}^{\rm X}(\epsilon,\bm x,\bm k)
&=&
\begin{pmatrix}
G^{\rm X}(\epsilon,\bm x,\bm k)&&F^{\rm X}(\epsilon,\bm x,\bm k)\\
\overline{F}^{\rm X}(\epsilon,\bm x,\bm k)&&\overline{G}^{{\rm X}}(\epsilon,\bm x,\bm k)
\end{pmatrix},
\end{eqnarray}
obeys the following left-hand Gor'kov equation,
\begin{eqnarray}
\label{left_Gorkov}
&&\left(\epsilon \check{\tau}_z - \check{\Delta}({\bm k})-\check{\sigma}_{\rm imp}\right)\check{\tau}_z\check{G}+\frac{i}{2}{\bm v}({\bm k})\cdot {\bm \nabla} \check{\tau}_z\check{G}-\xi_{\bm k}\check{\tau}_z\check{G}=1\,,
\end{eqnarray}
and the right-hand Gor'kov equation,
\begin{eqnarray}
\label{right_Gorkov}
&&\check{\tau}_z\check{G}\left(\epsilon \check{\tau}_z - \check{\Delta}({\bm k})-\check{\sigma}_{\rm imp}\right)-\frac{i}{2}{\bm v}({\bm k})\cdot {\bm \nabla} \check{\tau}_z\check{G}-\xi_{\bm k}\check{\tau}_z\check{G}=1\,.
\end{eqnarray}
Here, the superscript ${\rm X}={\rm R}, {\rm A}, {\rm K}$ represents the retarded, advanced and Keldysh components,
$\check{\tau}_{i}\; (i=x,y,z)$ are the Pauli matrices in the Nambu space,
$\xi_{\bm k}$ is the kinetic energy in the normal state relative to the chemical potential, and ${\bm v}({\bm k})={\bm \nabla}_{\bm k}\xi_{\bm k}$ is the quasiparticle velocity.
The superconducting order parameter matrix is $\check{\Delta}$, and $\check{\sigma}_{\rm imp}$ denotes the impurity self-energy.
Throughout this letter and supplemental material, we denote a $8\times8$ matrix in the Keldysh space as $\check{A}$, and a $4\times4$ matrix in the spin and the Nambu (particle-hole) space as $\underline{A}$.
If a matrix $\underline{A}$ is defined in the Nambu space, the corresponding matrix $\check{A}$ in the Keldysh space is,
\begin{eqnarray}
\check{A}=\underline{A}\otimes\openone=\begin{pmatrix}
\underline{A}&&0\\
0&&\underline{A}
\end{pmatrix}.
\end{eqnarray}

In the following we derive the quasiclassical transport equation.
We subtract Eq.~(\ref{right_Gorkov}) from Eq.~(\ref{left_Gorkov}) to obtain,
\begin{eqnarray}
\label{sub_Gorkov}
&&\left[\epsilon \check{\tau}_z - \check{\Delta}({\bm k})-\check{\sigma}_{\rm imp}, \check{\tau}_z\check{G}\right]
+i{\bm v}({\bm k})\cdot \check{\tau}_z{\bm \nabla} \check{G}=0.
\end{eqnarray}
It is convenient to define the quasiclassical Green's function~\cite{serene1983quasiclassical},
\begin{eqnarray}
\label{Greenfunc}
\check{g}(\epsilon,\bm x,{\bm k}_{\rm F})&\equiv&\int d\xi_{\bm k} \check{\tau}_z\check{G}(\epsilon,\bm x,\bm k)
\nonumber\\
&=&
\begin{pmatrix}
\underline{g}^{{\rm R}}(\epsilon,\bm x,{\bm k}_{\rm F})&&\underline{g}^{{\rm K}}(\epsilon,\bm x,{\bm k}_{\rm F})\\
0&&\underline{g}^{{\rm A}}(\epsilon,\bm x,{\bm k}_{\rm F})
\end{pmatrix},\\
\underline{g}^{\rm X}(\epsilon,\bm x,{\bm k}_{\rm F})&\equiv&\int d\xi_{\bm k} \underline{\tau}_z\underline{G}^{\rm X}(\epsilon,\bm x,\bm k).
\end{eqnarray}
In the quasiclassical limit, $(k_{\rm F}\xi)^{-1}\to 0$, we can assume the slow variation of the superconducting order parameter and the impurity self-energy with $\xi_{\bm k}$, relative to that of the Green's function.
In this limit, we express Eqs.~(\ref{sub_Gorkov}) as,
\begin{eqnarray}
\label{sub_Gorkov_1}
\left[\epsilon \check{\tau}_z - \check{\Delta}({\bm k}_{\rm F})-\check{\sigma}_{\rm imp}, \check{g}\right]+i{\bm v}_{\rm F}\cdot {\bm \nabla} \check{g}=0,
\end{eqnarray}
where now all of the momenta in Eqs.~(\ref{sub_Gorkov_1}) are fixed on the Fermi surface.
The quasiclassical Green's function is supplemented by the normalization condition, $\check{g}^2=-\pi^2$~\cite{eilenberger1968transformation}.
Eq.~(\ref{sub_Gorkov_1}) is the quasiclassical transport equation in the quasiclassical limit $(k_{\rm F}\xi)^{-1}\to 0$, which is known as the Eilenberger equation.

\section{Transport theory for spin-triplet superconductors at quasiclassical limit}
As discussed above, the transport properties of the superconductors is described by the quasiclassical Green's funcion.
For a triplet superconductor described by the $d$-vector ${\bm d}({\bm k})$, the superconducting gap matrix is given by,
\begin{eqnarray}
\check{\Delta}({\bm k}_{\rm F})&=&
\begin{pmatrix}
\underline{\Delta}({\bm k}_{\rm F})&&\underline{0}\\
\underline{0}&&\underline{\Delta}({\bm k}_{\rm F})
\end{pmatrix},\\
\underline{\Delta}({\bm k}_{\rm F})&=&
\begin{pmatrix}
0&&i({\bm \sigma}\cdot {\bm d}({{\bm k}_{\rm F}}))\sigma_y\\
i\sigma_y({\bm \sigma}\cdot {\bm d}^*({{\bm k}_{\rm F}}))&&0
\end{pmatrix}.
\end{eqnarray}
In this supplemental materials, we focus on the time-reversal invariant, unitary states, which satisfy,
\begin{eqnarray}
\underline{\Delta}^2({\bm k}_{\rm F})=-|{\bm d}({{\bm k}_{\rm F}})|^2\,.
\end{eqnarray}

In the quasiclassical transport theory, the temperature gradient can be incorporated through the spatial gradient~\cite{graf1996electronic,vorontsov2007unconventional}.
To consider response to the temperature gradient, we assume a local equilibrium $T=T({\bm x})$ and expand the spatial gradient as ${\bm \nabla} \to ({\bm \nabla} T) \frac{\partial}{\partial T}$.
Then, we obtain
\begin{eqnarray}
\label{Keldysh eq_w/_delT}
\left[ \epsilon\check{\tau}_z-\check{\Delta}-\check{\sigma}_{\rm imp},\check{g} \right]+(i{\bm v_{\rm F}}\cdot {\bm \nabla} T) \frac{\partial }{\partial T}\check{g}=0.
\end{eqnarray}

\section{Impurity self-energy}
Here, we explain the treatment of impurity scattering of quasiparticles and derive the $T$-matrix equation (Eq.~(3) in the main text).
The SNE in TRI TSCs arises from the asymmetric impurity scattering of quasiparticles mediated by the helical superconducting order.
It is necessary to include the effect of the multiple impurity scattering into the impurity self-energy for this effect.
For this purpose, we adopt the self-consistent $T$-matrix approximation (SCTA) to evaluate the impurity self-energy.

\begin{figure}[b]
\centering
\includegraphics[width=90mm]{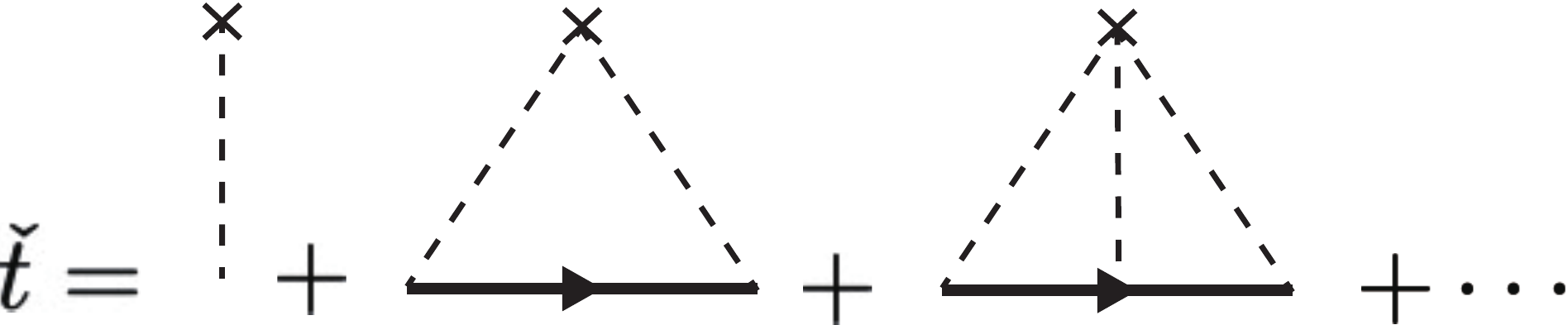}
\caption{Feynman diagrams for the $T$ matrix equation with a single impurity.
The X, the dashed lines, and the arrows are the impurity, the impurity potentials, and the Green's function, respectively.
}
\label{SCTA}
\end{figure}

The SCTA considers all of the non-crossing diagrams and then is obtained by imposing the self-consistency condition (Fig.~\ref{SCTA}).
The diagrams with the intersections of lines of the impurity potential are suppressed by the factor, $1/(k_{\rm F}\ell)$, where $\ell$ is the mean free path.
This factor is small in the quasicassical theory, allowing us to focus on the non-crossing diagrams.
First, we consider the case with a single impurity.
We assume the $\delta$-function impurity potential with the impurity potential strength, $V_{\rm imp}$.
For the SCTA, the $T$-matrix, $\check{t}$, for a single impurity is computed with the $T$-matrix equation,
\begin{eqnarray}
\label{T0}
\check{t}(\epsilon)=V_{\rm imp}+N(\epsilon_{\rm F})V_{\rm imp}\braket{\check{g}_0(\epsilon,{\bm k}_{\rm F})}_{{\rm FS}}\check{t}(\epsilon).
\end{eqnarray}
where $\check{g}_0(\epsilon,{\bm k}_{\rm F})$ is the quasiclassical Green's function in the clean system.
The Feynman diagrams for the $T$-matrix equation are shown in Fig.~\ref{SCTA}.
The $T$-matrix precisely accounts for the multiple scattering at the impurity site.

To consider the ensemble of impurities in real systems, we postulate the random distribution of impurity sites and take the statistical average :
\begin{eqnarray}
\label{selfen0}
\check{\sigma}_{\rm imp}(\epsilon)=n_{\rm imp}\check{t}(\epsilon),
\end{eqnarray}
where $n_{\rm imp}$ is the density of impurities.
The $T$-matrix, \eqref{selfen0}, describes the multiple scattering from the single impurity site.
This treatment of the impurity self-energy is called the $T$-matrix approximation.
The SCTA is obtained from the $T$-matrix approximation by assuming the self-consistency of the impurity self-energy.
This procedure is accomplished by modifying the $T$-matrix, $\check{t}$.
Replacing the Green's function in Eq.~\eqref{T0} by the dressed Green's function with the impurity self-energy, $\check{g}(\epsilon,{\bm k}_{\rm F})$, we obtain the impurity self-energy in the SCTA,
\begin{eqnarray}
\check{t}(\epsilon)=V_{\rm imp}+N(\epsilon_{\rm F})V_{\rm imp}\braket{\check{g}(\epsilon,{\bm k}_{\rm F})}_{{\rm FS}}\check{t}(\epsilon).
\end{eqnarray}
The SCTA describes the multiple scattering of quasiparticles which are already scattered by other impurities.

It is convenient to introduce the normal-state scattering rate, $\Gamma_{\rm imp}=\frac{n_{\rm imp}}{\pi N(\epsilon_{\rm F})}$, and the normal-state scattering phase shift, $\cot\delta = -1/[\pi N(\epsilon_{\rm F})V_{\rm imp}]$.
These quantities in the normal state are treated as parameters for calculations.
The normal state scattering rate has the dimension of the energy and should be set as $\Gamma_{\rm imp} \ll T_{\rm c}$ compatible with odd-parity superconductivity.
The limit $\delta \to 0$ corresponds to the weak impurity potential $V_{\rm imp}\to 0$ and is thus referred to as the Born limit.
The opposite limit $\delta \to \pi/2$ describes the strong impurity potential $V_{\rm imp}\to \infty$ and is referred to as the unitarity limit.
Using these normal state quantities, we recast Eq.~\eqref{selfen0} into Eq.~(3) in the main text,
\begin{eqnarray}
\label{selfen}
\check{\sigma}_{\rm imp}(\epsilon)=-\Gamma_{\rm imp} \left(\cot \delta+\left\langle\frac{ \check{g}(\epsilon,{\bm k}_{\rm F})}{\pi}\right\rangle_{{\rm FS}} \right)^{-1}.
\end{eqnarray}

\section{Spin current}
Next, we derive the expression of the spin current with the quasiclassical Green's function.
\begin{eqnarray}
\label{SC_Gorkov}
{\bm J}^{\sigma_\mu}(\bm x)&=&\int \frac{d\epsilon}{4\pi i}\int \frac{d{\bm k}}{(2\pi)^3}
{\rm Tr}_2\left[{\bm v}(\bm k) \sigma_\mu G^{{\rm K}}(\epsilon,\bm x, \bm k)\right] \;\;\;(\mu=x,y,z),
\end{eqnarray}
where $\sigma_\mu\;(\mu=x,y,z)$ represents the $2\times 2$ spin operator and ${\rm Tr}_2[\cdots]$ does the trace in the spin space.
Using the symmetry of the Keldysh Green's function~\cite{serene1983quasiclassical},
\begin{eqnarray}
G^{\rm K,\rm tr}(\epsilon,\bm x, \bm k)=-\overline{G}^{\rm K}(-\epsilon,\bm x,-\bm k),
\end{eqnarray}
we transform the spin current into,
\begin{eqnarray}
\label{ex_SC}
{\bm J}^{\sigma_{\mu}}(\bm x)&=&\int \frac{d\epsilon}{4\pi i}\int \frac{d{\bm k}}{(2\pi)^3}
{\rm Tr}_2\left[{\bm v}(\bm k) \sigma_{\mu} G^{{\rm K}}(\epsilon,\bm x, \bm k)\right]\nonumber\\
&=&\frac{1}{2}\int \frac{d\epsilon}{4\pi i}\int \frac{d{\bm k}}{(2\pi)^3}
{\rm Tr}_2\left[{\bm v}(\bm k) \sigma_{\mu} G^{{\rm K}}(\epsilon,\bm x, \bm k)\right]
-\frac{1}{2}\int \frac{d\epsilon}{4\pi i}\int \frac{d{\bm k}}{(2\pi)^3}
{\rm Tr}_2\left[{\bm v}(\bm k) \sigma_{\mu} \overline{G}^{{\rm K},{\rm tr}}(-\epsilon,\bm x, -\bm k)\right]\nonumber\\
&=&\frac{1}{2}\int \frac{d\epsilon}{4\pi i}\int \frac{d{\bm k}}{(2\pi)^3}
{\rm Tr}_2\left[{\bm v}(\bm k) \sigma_{\mu} G^{{\rm K}}(\epsilon,\bm x, \bm k)\right]
+\frac{1}{2}\int \frac{d\epsilon}{4\pi i}\int \frac{d{\bm k}}{(2\pi)^3}
{\rm Tr}_2\left[{\bm v}(\bm k)\sigma_{\mu} \overline{G}^{{\rm K},{\rm tr}}(\epsilon,\bm x, \bm k)\right]\nonumber\\
&=&\frac{1}{2}\int \frac{d\epsilon}{4\pi i}\int \frac{d{\bm k}}{(2\pi)^3}
{\rm Tr}_2\left[{\bm v}(\bm k) \sigma_{\mu} G^{{\rm K}}(\epsilon,\bm x, \bm k)\right]
+\frac{1}{2}\int \frac{d\epsilon}{4\pi i}\int \frac{d{\bm k}}{(2\pi)^3}
{\rm Tr}_2\left[{\bm v}(\bm k)\sigma_{\mu}^{\rm tr} \overline{G}^{{\rm K}}(\epsilon,\bm x, \bm k)\right]\nonumber\\
&=&\frac{1}{2}\int \frac{d\epsilon}{4\pi i}\int \frac{d{\bm k}}{(2\pi)^3}
{\rm Tr}\left[{\bm v}(\bm k) \underline{\sigma}_{\mu} \underline{G}^{{\rm K}}(\epsilon,\bm x, \bm k)\right],
\end{eqnarray}
where the $4\times 4$ spin operator,
\begin{eqnarray}
\underline{\sigma}_{\mu}=
\begin{pmatrix}
\sigma_{\mu}&&0\\
0&&\sigma_{\mu}^{\rm tr}
\end{pmatrix},
\end{eqnarray}
and the trace in the Nambu space ${\rm Tr}[\cdots]$ are introduced.
Using the definitions of the quasiclassical Green's function, we express Eq.~(\ref{ex_SC}) as,
\begin{eqnarray}
\label{ex_SC_1}
{\bm J}^{\sigma_{\mu}}(\bm x)&=&\frac{1}{2}N(\epsilon_{\rm F}) \int \frac{d\epsilon}{4\pi i} \left\langle{\rm Tr}\left[{\bm v_{\rm F}} \underline{\sigma}_{\mu} \underline{\tau}_z \underline{g}_{(0)}^{{\rm K}}(\epsilon,\bm x, {\bm k}_{\rm F})\right]\right\rangle_{\rm FS}+\mathcal{O}((k_{\rm F}\xi)^{-1}).
\end{eqnarray}
Here, $\braket{\cdots}_{\rm FS}$ is the normalized Fermi surface average ($\braket{1}_{\rm FS}=1$), $N(\epsilon)$ is the normal state density of states (DOS).

\section{Derivation of non-equilibrium Green's function}
For the determination of the SNE in TRS TSCs, we solve Eq.~(\ref{Keldysh eq_w/_delT}).
Let us consider the deviation $\delta \check{x}\;(x=g, \Delta, \sigma_{\rm imp})$ from the equilibrium function $\check{x}_{\rm eq}\;(x=g, \Delta, \sigma_{\rm imp})$.
As will be seen later, the correction to the superconducting gap matrix vanishes, and thus we set $\delta \check{\Delta}=0$~\cite{yip_ATHE}.

We first derive the equilibrium quasiclassical Green's function, which obeys,
\begin{eqnarray}
\label{Keldysh eq_eq}
\left[ \epsilon\check{\tau}_z-\check{\Delta}_{\rm eq}-\check{\sigma}_{\rm imp, eq},\check{g}_{\rm eq} \right]=0,
\end{eqnarray}
and is normalized by $\check{g}_{\rm eq}^2=-\pi^2$.
With this normalization, Eq.~(\ref{Keldysh eq_eq}) is easily solved to give,
\begin{eqnarray}
\label{g_eq_RA}
\underline{g}_{\rm eq}^{{\rm X}}&=&-\pi \frac{\underline{M}^{{\rm X}}}{D^{\rm X}}\;\;\; {\rm for}\;{\rm X}={\rm R},\;{\rm A},\\
\label{g_eq_K}
\underline{g}_{\rm eq}^{{\rm K}}&=&\left( \underline{g}_{\rm eq}^{{\rm R}}-\underline{g}_{\rm eq}^{{\rm A}} \right)\tanh \left(\frac{\epsilon}{2T}\right),
\end{eqnarray}
where $\underline{M}^{{\rm X}}=\tilde{\epsilon}^{\rm X}\underline{\tau}_z-\underline{\Delta}_{\rm eq}$, $D^{\rm X}=\sqrt{|{\bm d}_{\rm eq}({{\bm k}_{\rm F}})|^2-\tilde{\epsilon}^{{\rm X}\;2}}$ and $\tilde{\epsilon}^{\rm X}=\epsilon-\frac{1}{4}{\rm Tr}(\underline{\tau}_z\underline{\sigma}_{\rm eq, imp}^{\rm X})$.
The equilibrium impurity self-energy is diagonal in the Nambu space and independent of spins because the gap function averages to zero over the Fermi surface, $\braket{\underline{\Delta}_{\rm eq}({\bm k}_{\rm F})}_{\rm FS}=0$.
Note that the kinetic energy we consider is spin-independent.
We now determine the linear, in the thermal gradient, correction to the Green's function due to deviation from equilibrium.
This first-order correction of the retarded/advanced Green's function $\delta \underline{g}^{\rm X}\;({\rm X}={\rm R},\;{\rm A})$ satisfies
\begin{eqnarray}
\label{RA eq_noneq}
\left[ \underline{M}^{{\rm X}}, \delta \underline{g}^{{\rm X}}\right]-\left[ \delta \underline{\sigma}_{\rm imp}^{{\rm X}}, \underline{g}^{{\rm X}}_{\rm eq}\right]
+(i{\bm v_{\rm F}}\cdot {\bm \nabla} T) \frac{\partial}{\partial T} \underline{g}^{{\rm X}}_{\rm eq}=0\,,
\end{eqnarray}
and is normalized by $\{\underline{g}^{{\rm X}}_{\rm eq}, \delta \underline{g}^{{\rm X}}\}$=0.
With this normalization, we obtain
\begin{eqnarray}
\label{del g^X}
\delta \underline{g}^{{\rm X}}=\frac{\underline{g}^{{\rm X}}_{\rm eq}}{2\pi D^{\rm X}}\left( \left[\delta \underline{\sigma}^{{\rm X}}_{\rm imp},\underline{g}^{{\rm X}}_{\rm eq} \right]-(i{\bm v_{\rm F}}\cdot {\bm \nabla} T) \frac{\partial}{\partial T} \underline{g}^{{\rm X}}_{\rm eq} \right).
\end{eqnarray}
The second term of Eq.~(\ref{del g^X}) stems from the temperature dependence of the superconducting gap function, which is negligible at low-temperatures, $T/T_c\ll 1$.

The first-order correction to the Keldysh Green's function obeys the following transport equation,
\begin{eqnarray}
\label{K_eq_noneq}
&\left(\underline{M}^{{\rm R}} \delta \underline{g}^{{\rm K}}-\delta \underline{g}^{{\rm K}}\underline{M}^{{\rm A}}\right)-\left(\sigma_{\rm imp,eq0}^{\rm R}-\sigma_{\rm imp,eq0}^{\rm A}\right)\delta \underline{g}^{\rm K}
-\left( \underline{\sigma}_{\rm imp,eq}^{{\rm K}}\delta \underline{g}^{{\rm A}}- \delta\underline{g}^{{\rm R}}\underline{\sigma}_{\rm imp,eq}^{{\rm K}}\right)\nonumber\\
&-\left(\delta \underline{\sigma}_{\rm imp}^{{\rm R}} \underline{g}_{\rm eq}^{{\rm K}}-\underline{g}_{\rm eq}^{{\rm K}}\delta \underline{\sigma}_{\rm imp}^{{\rm A}}\right)-\left( \delta \underline{\sigma}_{\rm imp}^{{\rm K}}\underline{g}_{\rm eq}^{{\rm A}}- \underline{g}_{\rm eq}^{{\rm R}}\delta \underline{\sigma}_{\rm imp}^{{\rm K}}\right)+(i{\bm v_{\rm F}}\cdot {\bm \nabla} T)\frac{\partial}{\partial T} \underline{g}_{\rm eq}^{{\rm K}}=0.
\end{eqnarray}
where $\sigma_{\rm imp,eq0}^{{\rm R},{\rm A}}={\rm Tr}(\sigma_{\rm imp,eq}^{{\rm R},{\rm A}})$.
To solve Eq.~(\ref{K_eq_noneq}), it is convenient to define the anomalous Keldysh Green's function $\delta \underline{g}^{a}$ and the anomalous Keldysh impurity self-energy $\delta \underline{\sigma}_{\rm imp}^{a}$,
\begin{eqnarray}
\label{ga_K}
\delta \underline{g}^{a}&=&
\delta \underline{g}^{{\rm K}}-\left(\delta \underline{g}^{{\rm R}}-\delta \underline{g}^{{\rm A}}\right)
\tanh \left(\frac{\epsilon}{2T}\right),\\
\label{siga_K}
\delta \underline{\sigma}_{\rm imp}^{a}&=&\delta \underline{\sigma}_{\rm imp}^{{\rm K}}-\left(\delta \underline{\sigma}_{\rm imp}^{{\rm R}}-\delta \underline{\sigma}_{\rm imp}^{{\rm A}}\right)
\tanh \left(\frac{\epsilon}{2T}\right).
\end{eqnarray}
The anomalous Keldysh impurity self-energy is calculated from the $T$-matrix equation (\ref{selfen}).
Using Eqs.~(\ref{ga_K}-\ref{siga_K}), we obtain for this component,
\begin{eqnarray}
\label{T_mat_siga}
\delta \underline{\sigma}_{\rm imp}^{a}=\Gamma_{\rm imp}\left(\cot \delta +\left\langle \frac{\underline{g}_{\rm eq}^{\rm R}}{\pi} \right\rangle_{\rm FS}\right)^{-1}
\left\langle \frac{\delta \underline{g}^{a}}{\pi} \right\rangle_{\rm FS}\left(\cot \delta +\left\langle \frac{\underline{g}_{\rm eq}^{\rm A}}{\pi}\right\rangle_{\rm FS}\right)^{-1}\,.
\end{eqnarray}
Using these anomalous Keldysh functions, we transform Eq.~(\ref{K_eq_noneq}) into
\begin{eqnarray}
\label{aK eq_noneq}\left(\underline{M}^{{\rm R}} \delta \underline{g}^{a}-\delta \underline{g}^{a}\underline{M}^{{\rm A}}\right)-\left( \sigma_{\rm imp,eq0}^{\rm R}-\sigma_{\rm imp,eq0}^{\rm A} \right)\delta\underline{g}^{a}
+\left(\underline{g}_{\rm eq}^{{\rm R}}\delta \underline{\sigma}_{\rm imp}^{a}- \delta \underline{\sigma}_{\rm imp}^{a}\underline{g}_{\rm eq}^{{\rm A}} \right)-\frac{i\left(\epsilon {\bm v_{\rm F}}\cdot {\bm \nabla} T\right)}{2T^2\cosh^2 \left(\frac{\epsilon}{2T}\right)}
\left( \underline{g}_{\rm eq}^{{\rm R}}-\underline{g}_{\rm eq}^{{\rm A}}\right)=0.\nonumber\\
\end{eqnarray}
$\delta \underline{g}^{a}$ is normalized by $\underline{g}_{\rm eq}^{{\rm R}}\delta \underline{g}^{a}+\delta \underline{g}^{a}\underline{g}_{\rm eq}^{{\rm A}}=0$.
With this condition, we obtain the anomalous Keldysh Green's function,
\begin{eqnarray}
\label{gEli}
\delta \underline{g}^{a}&=&\delta \underline{g}^{a}_{\rm ns}+\delta \underline{g}^{a}_{\rm vc},\\
\label{gEli_ns}
\delta \underline{g}^{a}_{\rm ns}&=&\underline{N}^{\rm R}_{\rm eq}\left( \underline{g}_{\rm eq}^{{\rm R}}-\underline{g}_{\rm eq}^{{\rm A}}\right)\left(-\frac{i\left({\epsilon \bm v_{\rm F}}\cdot {\bm \nabla} T\right)}{2T^2\cosh^2\left(\frac{\epsilon}{2T}\right)}\right),\\
\label{gEli_vc}
\delta \underline{g}^{a}_{\rm vc}&=&
\underline{N}^{\rm R}_{\rm eq}\left(\underline{g}_{\rm eq}^{{\rm R}} \delta \underline{\sigma}_{\rm imp}^{a}- \delta \underline{\sigma}_{\rm imp}^{a}\underline{g}_{\rm eq}^{{\rm A}} \right),
\end{eqnarray}
where
\begin{eqnarray}
\label{NR}
\underline{N}^{\rm R}_{\rm eq}=\frac{\left(D^{\rm R}+D^{\rm A}\right)\left(-\frac{\underline{g}^{\rm R}_{\rm eq}}{\pi}\right)+\sigma_{\rm imp,eq0}^{\rm R}-\sigma_{\rm imp,eq0}^{\rm A} }{\left(D^{\rm R}+D^{\rm A} \right)^2+\left( \sigma_{\rm imp,eq0}^{\rm R}-\sigma_{\rm imp,eq0}^{\rm A} \right)^2}.
\end{eqnarray}
It is important to note that
$\delta \underline{g}^{a}_{\rm ns}$ depends only on the equilibrium impurity self-energies, but not on the anomalous component, $\delta \underline{\sigma}_{\rm imp}^a$, whereas $\delta \underline{g}^{a}_{\rm vc}$ is proportional to that component.
We emphasize that $\delta \underline{\sigma}_{\rm imp}^a$ corresponds to the vertex corrections in the diagrammatic calculation, which are critical for obtaining non-vanishing anomalous transport coefficients, see in next section.
Thus, we refer to $\delta \underline{g}^{a}_{\rm ns}$ as a non-self-consistent contribution, and $\delta \underline{g}^{a}_{\rm vc}$ as a vertex correction contribution to the anomalous Keldysh Green's function.

\section{Low-energy expansion for the equilibrium Green's function and the density of states in Dirac superconductors}
In this section, we derive the low-temperature formula for the SNE.
As we have seen earlier, the non-equilibrium Keldysh Green's function is proportional to ${\rm sech}^2(\epsilon/2T)$, which introduces a frequency cut-off $\epsilon \sim T$.
This frequency cut-off enables us to expand the non-equilibrium Green's function in $\epsilon$, which is referred to as the low-temperature expansion.
Eq.~(\ref{ex_SC_1}) shows that the non-equiliblium Green's function has to be expanded up to  the second-order in $\epsilon$ to obtain the finite spin current.
The second-order terms originate from (i) the $\epsilon^2$-dependence of the anomalous Keldysh impurity self-energy $\delta \sigma_{\rm imp}^a$, and (ii) the combination of the $\epsilon$-linear dependence of the equilibrium Green's function $\check{g}_{\rm eq}$ and the anomalous Keldysh impurity self-energy $\delta \sigma_{\rm imp}^a$.
To obtain the low-temperature formula for the SNE, we perform the low-energy expansion for the equilibrium Green's function.
The quasiparticle DOS is also expanded in order to understand the impurity potential dependence in the SNE.

We introduce the following notations for the expansion, starting with the impurity self-energy,
\begin{eqnarray}
\underline{\sigma}_{\rm eq, imp}^{\rm R}&=&\sigma_{0,{\rm imp}}^{\rm R}(\epsilon)+\sigma_{z,{\rm imp}}^{\rm R}(\epsilon)\underline{\tau}_z,\\
\sigma_{0,{\rm imp}}^{\rm R}(\epsilon)&=&\sum_{n=0}^{\infty}\sigma_{0,{\rm imp}}^{(n){\rm R}}\epsilon^n\,,\\
\sigma_{z,{\rm imp}}^{\rm R}(\epsilon)&=&\sum_{n=0}^{\infty}\sigma_{z,{\rm imp}}^{(n){\rm R}}\epsilon^n\,.
\end{eqnarray}
Similarly, the Fermi-surface average of the equilibrium Green's function is written as,
\begin{eqnarray}
\left\langle\frac{\underline{g}_{\rm eq}^{\rm R}}{\pi}\right\rangle_{\rm FS}&=&-\left\langle\frac{\underline{M}^{\rm R}}{D^{\rm R}}\right\rangle_{\rm FS}=\sum_{n=0}^{\infty}G^{(n){\rm R}}\epsilon^n\underline{\tau}_z\,,
\end{eqnarray}
where we used $\braket{\underline{\Delta}_{\rm eq}({\bm k}_{\rm F})}_{\rm FS}=0$ for spin-triplet superconductors.
With this momentum average of the Green's function, the expansion of the quasiparticle DOS is,
\begin{eqnarray}
N_s(\epsilon)&\equiv&-\frac{N(\epsilon_{\rm F})}{4}{\rm Im}\left({\rm Tr} \left[\underline{\tau_z}\left\langle\frac{\underline{g}_{\rm eq}^{\rm R}}{\pi}\right\rangle_{\rm FS}\right]\right)\nonumber\\
&=&\sum_{n=0}^{\infty}N_s^{(n)}\epsilon^n=-N(\epsilon_{\rm F})\sum_{n=0}^\infty {\rm Im}G^{(n){\rm R}}\epsilon^n.
\end{eqnarray}
The impurity self-energy is calculated from the $T$-matrix equation (\ref{selfen}),
\begin{eqnarray}
\sigma_{0,{\rm imp}}^{\rm R}(\epsilon)&=&-\Gamma_{\rm imp}\frac{\cot \delta}{\cot^2 \delta -(\sum_n G^{(n)\rm R}\epsilon^n)^2 },\\
\sigma_{z,{\rm imp}}^{\rm R}(\epsilon)&=&\Gamma_{\rm imp}\frac{\sum_n G^{(n)\rm R}\epsilon^n}{\cot^2 \delta -(\sum_n G^{(n)\rm R}\epsilon^n)^2 }.
\end{eqnarray}
From these equations, the coefficients $\{\sigma_{0,{\rm imp}}^{(n) \rm R}\}$ and $\{\sigma_{z,{\rm imp}}^{(n) \rm R}\}$ are given by,
\begin{eqnarray}
\label{sigma00}
\sigma_{0,{\rm imp}}^{(0) \rm R}&=&-\frac{\Gamma_{\rm imp}\cot \delta}{\cot^2 \delta -G^{(0){\rm R}\; 2}},\\
\sigma_{0,{\rm imp}}^{(1) \rm R}&=&-\Gamma_{\rm imp}\cot \delta\frac{2G^{(0)\rm R}G^{(1)\rm R}}{(\cot^2 \delta -G^{(0){\rm R}\; 2})^2},\\
\sigma_{0,{\rm imp}}^{(2) \rm R}&=&-\Gamma_{\rm imp}\cot \delta\left[\frac{(G^{(1){\rm R}\; 2}+2G^{(0)\rm R }G^{(2)\rm R })}{(\cot^2 \delta -G^{(0){\rm R}\; 2})^2}+\frac{4G^{(0){\rm R} 2}G^{(1){\rm R}\; 2}}{(\cot^2 \delta -G^{(0){\rm R}\; 2})^3}\right],\\
\sigma_{z,{\rm imp}}^{(0) \rm R}&=&\frac{\Gamma_{\rm imp}G^{{(0)}{\rm R}}}{\cot^2 \delta -G^{(0){\rm R}\; 2}},\\
\sigma_{z,{\rm imp}}^{(1) \rm R}&=&\Gamma_{\rm imp}\frac{(\cot^2 \delta+G^{(0){\rm R}\; 2})G^{(1)\rm R}}{(\cot^2 \delta -G^{(0){\rm R}\; 2})^2},\\
\sigma_{z,{\rm imp}}^{(2) \rm R}&=&\Gamma_{\rm imp}\left[\frac{3G^{(0)\rm R }G^{(1){\rm R}\; 2}}{(\cot^2 \delta -G^{(0){\rm R}\; 2})^2}+\frac{(\cot^2 \delta+G^{(0){\rm R}\; 2})G^{(2){\rm R}} }{(\cot^2 \delta -G^{(0){\rm R}\; 2})^2}+\frac{4G^{(0){\rm R} \; 3}G^{(1){\rm R}\; 2}}{(\cot^2 \delta -G^{(0){\rm R}\; 2})^3}\right].
\end{eqnarray}
From the equilibrium Green's function (\ref{g_eq_RA}), the coefficients $\{ G^{(n){\rm R}} \}$ are given by,
\begin{eqnarray}
G^{(0){\rm R}}&=&\left\langle\frac{\sigma_{z,{\rm imp}}^{(0) \rm R}}{\sqrt{|{\bm d}_{\rm eq}({{\bm k}_{\rm F}})|^2-\sigma_{z,{\rm imp}}^{(0) {\rm R}\;2}}}\right\rangle_{\rm FS},\\
G^{(1){\rm R}}&=&-\left\langle\frac{|{\bm d}_{\rm eq}({{\bm k}_{\rm F}})|^2}{\left[|{\bm d}_{\rm eq}({{\bm k}_{\rm F}})|^2-\sigma_{z,{\rm imp}}^{(0) {\rm R}\;2}\right]^{\frac{3}{2}}}\right\rangle_{\rm FS}(1-\sigma_{z,{\rm imp}}^{(1) \rm R}),\\
\label{G2}
G^{(2){\rm R}}&=&\frac{3}{2}\left\langle\frac{|{\bm d}_{\rm eq}({{\bm k}_{\rm F}})|^2\sigma_{z,{\rm imp}}^{(0) \rm R}}{\left[|{\bm d}_{\rm eq}({{\bm k}_{\rm F}})|^2-\sigma_{z,{\rm imp}}^{(0) {\rm R}\; 2}\right]^{\frac{3}{2}}}\right\rangle_{\rm FS}(1-\sigma_{z,{\rm imp}}^{(1) \rm R})^2+\left\langle\frac{|{\bm d}_{\rm eq}({{\bm k}_{\rm F}})|^2}{\left[|{\bm d}_{\rm eq}({{\bm k}_{\rm F}})|^2-\sigma_{z,{\rm imp}}^{(0) {\rm R}\;2}\right]^{\frac{3}{2}}}\right\rangle_{\rm FS}\sigma_{z,{\rm imp}}^{(2) \rm R}.
\end{eqnarray}
Solving Eqs.~(\ref{sigma00}-\ref{G2}), we obtain the impurity self-energy,
\begin{eqnarray}
\label{low_en_eqselfen}
\underline{\sigma}_{\rm eq, imp}^{\rm R}&=&\left(-\frac{\Gamma_{\rm imp}\cot \delta}{\cot^2 \delta +n_s^2}-4iz^{-1}X\epsilon\right)+\left(-\frac{i\Gamma_{\rm imp}n_s}{\cot^2 \delta +n_s^2}-4z^{-1}Y\epsilon \right)\underline{\tau}_z+\mathcal{O}(\epsilon^2),
\end{eqnarray}
and the DOS,
\begin{eqnarray}
\label{N_s}
N_s(\epsilon)&=&N(\epsilon_{\rm F})\left[n_s+z^{-3}\left(\frac{3}{2}\gamma\left\langle \frac{|{\bm d}_{\rm eq}({{\bm k}_{\rm F}})|^2}{D_{\rm LT}^5}\right\rangle_{\rm FS}+\frac{4\left(Yn_s+X\cot \delta \right)}{\cot^2 \delta +n_s^2}\left\langle \frac{|{\bm d}_{\rm eq}({{\bm k}_{\rm F}})|^2}{D_{\rm LT}^3}\right\rangle_{\rm FS}\right)\epsilon^2\right]+\mathcal{O}(\epsilon^3),
\end{eqnarray}
where $D_{\rm LT}=\sqrt{|{\bm d}_{\rm eq}({{\bm k}_{\rm F}})|^2+\gamma^2}$, $\gamma \equiv \frac{i}{2}{\rm Tr}(\underline{\tau}_z\underline{\sigma}_{\rm imp,eq}^{\rm R}(0))$ and $n_s\equiv \frac{N_s(0)}{N(\epsilon_{\rm F})} = \left\langle \frac{\gamma}{D_{\rm LT}}\right\rangle_{\rm FS}$.
In the equations above, we introduced these dimensionless quantities,
\begin{eqnarray}
X&=& \frac{\Gamma_{\rm imp}n_s \cot \delta  }{2\left(\cot^2 \delta +n_s^2\right)^2 }\left\langle \frac{|{\bm d}_{\rm eq}({{\bm k}_{\rm F}})|^2}{D_{LT}^3}\right\rangle_{\rm FS},\\
Y&=& \frac{\Gamma_{\rm imp}\left(\cot^2 \delta -n_s^2\right) }{4 \left(\cot^2 \delta +n_s^2\right)^2 }\left\langle \frac{|{\bm d}_{\rm eq}({{\bm k}_{\rm F}})|^2}{D_{LT}^3}\right\rangle_{\rm FS}\,,
\end{eqnarray}
and the renormalization constant,
\begin{eqnarray}
z\equiv (1-\sigma_{z,\rm imp}^{(1)})^{-1}=1-4Y\,.
\end{eqnarray}

The factor $Y\propto \cot^2\delta -n_s^2$ sensitively depends on the character of the impurity bound states in unconventional superconductors.
In such superconductors, Andreev scattering at individual impurities creates resonant (or even truly bound) quasiparticle states, at energies that vary from mid-gap in the unitarity limit ($\delta=\pi/2$) to near the gap edge in the Born limit ($\delta\rightarrow 0$).
At finite concentration of impurities, these localized states form an impurity band, still peaked at the resonance energy, but with a finite width,  which significantly changes the quasiparticle DOS.	At unitarity, since each impurity state is at $\epsilon=0$, the impurity band has the peak of the DOS at this energy.
When the impurity potential is weaker, the tails of the band produce a finite DOS at $\epsilon=0$ for sufficient density of scatterers, but the peak remains at finite $\epsilon$. It follows that the low-energy quasiparticle DOS behaves as $N_s \simeq N_s^{(0)}+N_s^{(2)} \epsilon^2$, where $N_s^{(2)}\geq 0$ for weak scatterers, whereas $N_s^{(2)}<0$ in the vicinity of the unitarity limit.
We note that the factor $Y$ in Eq.~(\ref{N_s})  describes this competition, and takes a positive values for the weak impurity potential, while its sign becomes negative for near-unitarity scattteres.

The linear, in $\epsilon$, corrections appear in the real part of the equilibrium Green's function $\underline{g}^{\rm R,A}_{\rm eq}$,
\begin{eqnarray}
\label{geq_lowenR}
\underline{g}_{\rm eq,LT}^{\rm R}(\epsilon,{\bm k}_{\rm F})&=&-\pi \frac{A(\epsilon,{\bm k}_{\rm F})}{D_{\rm LT}({\bm k}_{\rm F})}\left[(z^{-1}\epsilon+i\gamma)\underline{\tau}_z-\underline{\Delta}_{\rm eq}({\bm k}_{\rm F})\right],\\
\label{geq_lowenA}
\underline{g}_{\rm eq,LT}^{\rm A}(\epsilon,{\bm k}_{\rm F})&=&-\pi \frac{A^*(\epsilon,{\bm k}_{\rm F})}{D_{\rm LT}({\bm k}_{\rm F})}\left[(z^{-1}\epsilon-i\gamma)\underline{\tau}_z-\underline{\Delta}_{\rm eq}({\bm k}_{\rm F})\right],
\end{eqnarray}
where $A(\epsilon,{\bm k}_{\rm F})=1+\frac{iz^{-1}\gamma}{D_{\rm LT}^2({\bm k}_{\rm F})}\epsilon$.

\section{Low-temperature expansion analysis in Dirac superconductors}
Now, we perform the low-temperature expansion analysis of the SNE in DSCs.
We assume the temperature gradient along the $y$-direction, the helical $p$-wave pairing ${\bm d}_{{\rm DSC},xy}(\bm k)=\Delta \left( k_x,k_y,0 \right)/k_{\rm F}$ on the spherical Fermi surface.
The helical $p$-wave paring gives rise to two Dirac points on the north and south poles on the Fermi sphere, and thus describes DSCs.
In the helical $p$-wave superconductor, the spin $\underline{\sigma}_z$ is conserved and the superconducting gap matrix $\underline{\Delta}({\bm k}_{\rm F})$ is block-diagonal, allowing us to perform the low-temperature expansion in each of the $\sigma_z$-subspaces.
In the $\sigma_z$-subspace, the quasiclassical Green's function $\check{g}^{\sigma_z}$ obeys, once again, the Eilenberger equation,
\begin{eqnarray}
\label{Keldysh eq_sz}
\left[ \epsilon\check{\tau}_z-\check{\Delta}^{\sigma_z}-\check{\sigma}_{\rm imp}^{\sigma_z},\check{g}^{\sigma_z} \right]+(i{\bm v_{\rm F}}\cdot {\bm \nabla} T) \frac{\partial}{\partial T} \check{g}^{\sigma_z}=0\,,
\end{eqnarray}
where $\check{\sigma}_{\rm imp}^{\sigma_z}$ is the impurity self-energy in the same spin subspace, and the gap function,
\begin{eqnarray}
\label{gap_mat_sz}
\underline{\Delta}^{\sigma_z}
&=&
\begin{pmatrix}
0&&-\sigma_z\Delta e^{-i\sigma_z \phi_k}\sin \theta_k\\
\sigma_z\Delta^* e^{i\sigma_z \phi_k}\sin \theta_k&&0
\end{pmatrix},
\end{eqnarray}
depends on the polar coordinates $(\theta_k, \phi_k)$ in the momentum space.
From now on, $\Delta \in \mathbb{R}$ is set to be real by the gauge transformation~\cite{ren_gap}.

To obtain all of the $\epsilon^2$-term in the anomalous Keldysh Green's function, the anomalous Keldysh impurity self-energy has to be expanded up to the second-order in $\epsilon$.
Here, we expand $\delta \underline{\sigma}_{\rm imp}^{\sigma_z,a}$ as,
\begin{eqnarray}
\delta \underline{\sigma}_{\rm imp,LT}^{\sigma_z a}(\epsilon)=\delta \underline{\sigma}_{\rm imp}^{\sigma_z (1)a}\epsilon+\delta \underline{\sigma}_{\rm imp}^{\sigma_z (2)a}\epsilon^2+\mathcal{O}(\epsilon^3).
\end{eqnarray}
Up to the first-order in $\epsilon$, Eq.~(\ref{NR}) gives,
\begin{eqnarray}
\label{NR_lowen}
\underline{N}^{\rm R}_{\rm eq,LT}(\epsilon)=-\frac{\underline{g}_{\rm eq,LT}^R}{2\pi D_{\rm LT}}-\frac{2iz^{-1}X\epsilon}{D_{\rm LT}^2}+\mathcal{O}(\epsilon^2).
\end{eqnarray}
Using the low-energy functions (\ref{geq_lowenR}), (\ref{geq_lowenA}) and (\ref{NR_lowen}), the anomalous Keldysh Green's function is given by
\begin{eqnarray}
\label{gEli_LT}
\delta \underline{g}^{\sigma_z,a}_{\rm LT}&=&\delta \underline{g}^{\sigma_z,a}_{\rm ns,LT}+\delta \underline{g}^{\sigma_z,a}_{\rm vc,LT},\\
\label{gEli_ns_LT}
\delta \underline{g}^{\sigma_z,a}_{\rm ns, LT}&=&
\underline{N}^{\rm R}_{\rm eq,LT}\left( \underline{g}_{\rm eq, LT}^{\sigma_z,\rm R}-\underline{g}_{\rm eq, LT}^{\sigma_z,{\rm A}}\right)\left(\frac{-i \left(\epsilon{\bm v_{\rm F}}\cdot {\bm \nabla} T\right)}{2 T^2 \cosh^2 \left(\frac{\epsilon}{2T}\right)}\right),\\
\label{gEli_vc_LT}
\delta \underline{g}^{\sigma_z,a}_{\rm vc, LT}&=&
\underline{N}^{\rm R}_{\rm eq,LT}
\left( \underline{g}_{\rm eq, LT}^{\sigma_z,\rm R} \delta \underline{\sigma}_{\rm imp,LT}^{\sigma_z,a}-\delta \underline{\sigma}_{\rm imp,LT}^{\sigma_z,a}\underline{g}_{\rm eq, LT}^{\sigma_z,{\rm A}}\right)\,.
\end{eqnarray}
From Eq.~(\ref{T_mat_siga}), we obtain the $T$-matrix equation for $\delta \underline{\sigma}_{\rm imp,LT}^{\sigma_z,a}$,
\begin{eqnarray}
\label{eq_selfen_noneq_LT_1}
\delta \underline{\sigma}^{\sigma_z,a}_{\rm imp,LT}&=&\Gamma_{\rm imp}
\left( \cot \delta +\left\langle \frac{\underline{g}_{\rm eq,LT}^{\sigma_z,{\rm R}}}{\pi}\right\rangle_{\rm FS} \right)^{-1} \left\langle \frac{\delta \underline{g}^{\sigma_z,a}_{\rm ns, LT}+\delta \underline{g}^{\sigma_z,a}_{\rm vc, LT}}{\pi} \right\rangle_{\rm FS}
\left( \cot \delta +\left\langle \frac{\underline{g}_{\rm eq,LT}^{\sigma_z,{\rm A}}}{\pi}\right\rangle_{\rm FS} \right)^{-1}.
\end{eqnarray}
We can evaluate $\left\langle \frac{\delta \underline{g}^{\sigma_z,a}_{\rm ns, LT}}{\pi} \right\rangle_{\rm FS}$ straightforwardly. Then, we obtain,
\begin{eqnarray}
\label{eq_selfen_noneq_LT_2}
&&\Gamma_{\rm imp}
\left( \cot \delta +\left\langle \frac{\underline{g}_{\rm eq,LT}^{\sigma_z {\rm R}}}{\pi}\right\rangle_{\rm FS} \right)^{-1} \left\langle \frac{\delta \underline{g}^{\sigma_z,a}_{\rm ns, LT}}{\pi} \right\rangle_{\rm FS}
\left( \cot \delta +\left\langle \frac{\underline{g}_{\rm eq,LT}^{\sigma_z, {\rm A}}}{\pi}\right\rangle_{\rm FS} \right)^{-1}\nonumber\\
&&=\left(X\underline{\tau}_x+Y\underline{\tau}_y\right)\frac{\gamma \epsilon v_{\rm F}\left(\partial_y T\right)}{\Delta_{\rm eq} T^2 \cosh^2 \left(\frac{\epsilon}{2T}\right)}+\mathcal{O}(\epsilon^3).
\end{eqnarray}
To solve Eq.~(\ref{eq_selfen_noneq_LT_1}) self-consistently, we assume the following form of the anomalous Keldysh impurity self-energy,
\begin{eqnarray}
\label{eq_selfen_noneq_LT_3}
\delta \underline{\sigma}_{\rm imp,LT}^{\sigma_z(n)a}=\left(\tilde{X}^{(n)}\underline{\tau}_x+\tilde{Y}^{(n)}\underline{\tau}_y\right)\frac{\gamma  v_{\rm F} \left(\partial_y T\right)}{\Delta_{\rm eq} T^2 \cosh^2 \left(\frac{\epsilon}{2T}\right)}.\;\;\;\;(n=1,2).
\end{eqnarray}
With this form of the impurity self-energy, we recast Eq.~(\ref{eq_selfen_noneq_LT_1}) into,
\begin{eqnarray}
\label{eq_selfen_noneq_LT_4}
\begin{pmatrix}
1-2Y&&-2X\\
2X&&1-2Y
\end{pmatrix}
\begin{pmatrix}
\tilde{X}^{(1)}\\
\tilde{Y}^{(1)}
\end{pmatrix}
=
\begin{pmatrix}
X\\
Y
\end{pmatrix}.
\end{eqnarray}
\begin{eqnarray}
\label{eq_selfen_noneq_LT_5}
\begin{pmatrix}
1-2Y&&-2X\\
2X&&1-2Y
\end{pmatrix}
\begin{pmatrix}
\tilde{X}^{(2)}\\
\tilde{Y}^{(2)}
\end{pmatrix}
=
\begin{pmatrix}
0\\
0
\end{pmatrix}.
\end{eqnarray}
From Eqs.~(\ref{eq_selfen_noneq_LT_4}-\ref{eq_selfen_noneq_LT_5}), we obtain the anomalous Keldysh impurity self-energy,
\begin{eqnarray}
\tilde{X}^{(1)}&=&\frac{X}{\rm Det},\\
\tilde{Y}^{(1)}&=&\frac{1}{\rm Det}\left[Y
-\frac{\Gamma_{\rm imp}^2  }{8\left(\cot^2 \delta +n_s^2\right)^2}\left\langle \frac{|{\bm d}_{\rm eq}({{\bm k}_{\rm F}})|^2}{D_{\rm LT}^3}\right\rangle_{\rm FS}^2\right],\\
\tilde{X}^{(2)}&=&\tilde{Y}^{(2)}=0,
\end{eqnarray}
where ${\rm Det}$ is the determinant of the matrix in Eq.~(\ref{eq_selfen_noneq_LT_4}),
\begin{eqnarray}
{\rm Det}&\equiv& 1-\frac{\Gamma_{\rm imp} \left(\cot^2 \delta -n_s^2\right)  }{\left(\cot^2 \delta +n_s^2\right)^2}\left\langle\frac{|{\bm d}_{\rm eq}({{\bm k}_{\rm F}})|^2}{D_{\rm LT}^3}\right\rangle_{\rm FS}
+\frac{\Gamma_{\rm imp}^2 }{4\left(\cot^2 \delta +n_s^2\right)^2}\left\langle\frac{|{\bm d}_{\rm eq}({{\bm k}_{\rm F}})|^2}{D_{\rm LT}^3}\right\rangle_{\rm FS}^2.
\end{eqnarray}

Substituting the anomalous impurity self-energy into Eq.~(\ref{gEli_LT}), we obtain the anomalous Keldysh Green's function.
Note that the obtained anomalous Keldysh Green's function at low-temperatures does not renormalize the superconducting gap function.
From the the anomalous Keldysh Green's function, which is due to the vertex corrections, we obtain the SNC,
\begin{eqnarray}
\label{spinNernst_coef}
\frac{\alpha^{\sigma_z}_{xy}}{N(\epsilon_{\rm F})v_{\rm F}^2 }&=&-\frac{z^{-1}\pi^2 T}{3|\Delta_{\rm eq}|^2{\rm Det}}\gamma \left\langle \frac{|{\bm d}_{\rm eq}({{\bm k}_{\rm F}})|^2}{D_{\rm LT}^3}\right\rangle_{\rm FS}\left(Y+4X^2-\frac{\Gamma_{\rm imp}^2  }{8\left(\cot^2 \delta +n_s^2\right)^2}\left\langle \frac{|{\bm d}_{\rm eq}({{\bm k}_{\rm F}})|^2}{D_{\rm LT}^3}\right\rangle_{\rm FS}^2\right)+\mathcal{O}(T^2).
\end{eqnarray}
The $Y$-term in the SNC originates from the frequency dependence of the $\underline{g}_{\rm eq,LT}^{\rm R}$ whereas the $X^2$-term in the stems from the frequency dependence of $\underline{N}^{\rm R}_{\rm eq,LT}$.
This low-energy formula for the SNC can reproduce the numerically calculated result, Fig.~2~(c) in the main text.
In the clean systems, Eq.~(\ref{spinNernst_coef}) reduces to Eq.~(6),
\begin{eqnarray}
\label{spinNernst_coef}
&&\frac{\alpha^{\sigma_z}_{xy}}{N(\epsilon_{\rm F})v_{\rm F}^2 }=
-\frac{\pi^2\gamma \Gamma_{\rm imp} T }{12|\Delta_{\rm eq}|^2}
\frac{\cot^2 \delta-{n_s^2}}{\left(\cot^2 \delta +n_s^2\right)^2 }\left\langle\frac{|{\bm d}_{\rm eq}({\bm k_{\rm F}})|^2}{D_{\rm LT}^3}\right\rangle_{\rm FS}^2+
\mathcal{O}(T^2,\Gamma_{\rm imp}^3)\,.
\end{eqnarray}

\section{Spin-Nernst effect in U$_{1-x}$Th$_x$Be$_{13}$}
Here we briefly discuss the most promising candidates for the order parameter in U$_{1-x}$Th$_x$Be$_{13}$, and connect those proposals with the symmetry of the spin-Nernst response.

\begin{figure}
\centering
\includegraphics[width=8cm]{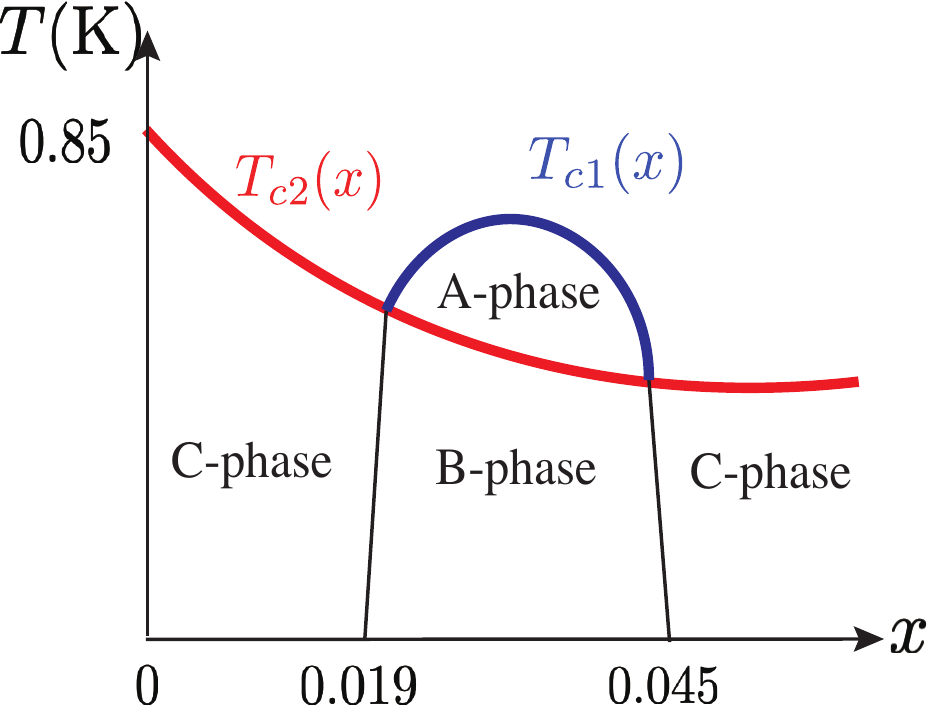}
\caption{
The multiple superconducting phases in U$_{1-x}$Th$_x$Be$_{13}$ in the $x$-$T$ plane.
}
\label{ube13_phase}
\end{figure}

U$_{1-x}$Th$_x$Be$_{13}$, discovered in 1985, is a spin-triplet superconductor exhibiting (at least) three-distinct superconducting phases depending on the temperature, $T$, and the dopant concentration, $x$, see Fig.~\ref{ube13_phase}. The parent material, UBe$_{13}$, undergoes a superconducting transition at  $T_{c2}(x=0)\sim 0.85$ K.
The transition temperature of this so-called C-phase decreases with increasing Th concentration. However, in
a narrow dopant region, $0.02\leq x\leq 0.04$, there exists a double superconducting transition, with transition temperatures, $T_{c1}(x)$ and $T_{c2}(x)$ in Fig.~\ref{ube13_phase}.
The high-temperature superconducting phase, $T_{c1}(x)(\geq T_{c2}(x))$ is denoted as the A-phase, while 
the low-temperature phase in this doping is distinct from the original C-phase, and is referred to as the B-phase. The pairing symmetry that gives rise to this phase diagram remains an unsolved issue.



Recent angle-resolved specific heat measurement revealed that the C-phase has a full-gap structure~\cite{shimizu2017quasiparticle}.
According to the $\mu$SR measurements, time-reversal symmetry is broken only in the low-temperature B-phase, and is preserved in the A and C phases~\cite{Heffner_UBe13}.

There are many candidates for the order parameter in this material, but the recent studies for the node and spin structures narrow down the possible pairing symmetry.
The leading scenarios are (i) the degenerate scenario, proposed by K. Machida recently~\cite{machida2018spin} and (ii) the accidental scenario, extensively studied by M.~Sigrist and T.~M.~Rice~\cite{sigrist_UBe13}.

The degenerate scenario postulates the existence of a higher (at least two)-dimensional irreducible representation for the order parameter symmetry that
allows for different the combination of the basis functions, naturally explaining the multiple superconducting phases.
Among such higher-order irreducible representations, the $E_u$ state consistently explains the node structure and the Knight-shift measurements.
The basis functions of the $E_u$ state are
\begin{eqnarray}
\hat{\bm l}_{E_{\rm u},1}(\hat{\bm k})&=&\sqrt{\frac{3}{2k_{\rm F}}}(k_x,-k_y,0),\\
\hat{\bm l}_{E_{\rm u},2}(\hat{\bm k})&=&\frac{1}{2k_{\rm F}}(-k_x,-k_y,2k_z).
\end{eqnarray}
We combine the basis functions to construct the candidates of the $d$-vector, ${\bm d}_\Gamma(\bm k)$, in the $\Gamma=A$, $B$ and $C$ phase as,
\begin{eqnarray}
{\bm d}_{\rm A}(\hat{\bm k})&=&\Delta_{E_{\rm u}}\hat{\bm l}_{E_{\rm u},1}(\hat{\bm k}),\\
{\bm d}_{\rm B}(\hat{\bm k})&=&\Delta_{E_{\rm u}}\left(\hat{\bm l}_{E_{\rm u},1}(\hat{\bm k})+i\hat{\bm l}_{E_{\rm u},2}(\hat{\bm k})\right),\\
{\bm d}_{\rm C}(\hat{\bm k})&=&\Delta_{E_{\rm u}}\hat{\bm l}_{E_{\rm u},2}(\hat{\bm k}),
\end{eqnarray}
where $\Delta_{E_{\rm u}}$ is the order parameter amplitude for the $E_u$ state.
The $E_u$ state predicts the biaxial nematic state in the A-phase and the uniaxial nematic state in the C-phase.
Note that the biaxial nematic state in the A-phase is the same as the model of DSCs discussed in the main text.
The C-phase is also similar to the BW state but with the anisotropy of the SNC, realizing TRI TSCs.
The time-reversal symmetry broken state described by ${\bm d}_{\rm B}(\hat{\bm k})\propto (k_x, \omega k_y, \omega^2k_z)\; (\omega^3=1)$ is referred to as the cyclic $p$-wave state and its topological properties was recently investigated.
The cyclic $p$-wave state has the Weyl node in the (111) direction and its equivalent directions, realizing Weyl superconductors~\cite{mizushima2018topology}.

From the similarity (or equivalence) between the $E_u$ state and our models, the A-phase supported by the $E_u$ state exhibits a finite spin-Nernst signal, $\alpha^{\sigma_z}_{xy}$.
The C-phase realized in the $E_u$ state manifests finite spin-Nernst tensor coefficients, $\alpha^{\sigma_z}_{xy},\; \alpha^{\sigma_x}_{yz}\; \alpha^{\sigma_y}_{zx}$.
Due to the nematicity of the $E_u$ state, these SNCs becomes anisotropic: $\alpha^{\sigma_z}_{xy}\neq \alpha^{\sigma_x}_{yz}= \alpha^{\sigma_y}_{zx}$.

The accidental scenario relies on an accidental degeneracy of the transition temperatures in different irreducible representations.
The most plausible choice of the order parameters is the $A_{1u}$ state in the C-phase, the $A_{2u}$ state in the A-phase, and $A_{1u}+i A_{2u}$ state in the B-phase.
With the use of the basis functions of the $A_{1u}$ and $A_{2u}$ states,
\begin{eqnarray}
\hat{\bm l}_{{\rm A}_{1u}}(\hat{\bm k})&=&\frac{1}{k_{\rm F}}(k_x,k_y,k_z),\\
\hat{\bm l}_{{\rm A}_{2u}}(\hat{\bm k})&=&\frac{\sqrt{35}}{2k_{\rm F}^3}(k_x(k_y^2-k_z^2),k_y(k_z^2-k_x^2),k_z(k_x^2-k_y^2)),
\end{eqnarray}
we obtain the possible order parameters in each superconducting phase,
\begin{eqnarray}
{\bm d}_{\rm A}(\hat{\bm k})&=&\Delta_{{\rm A}_{2u}}\hat{\bm l}_{{\rm A}_{2u}}(\hat{\bm k}),\\
{\bm d}_{\rm B}(\hat{\bm k})&=&\Delta_{{\rm A}_{1u}}\hat{\bm l}_{{\rm A}_{1u}}(\hat{\bm k})+i\Delta_{{\rm A}_{2u}}\hat{\bm l}_{{\rm A}_{2u}}(\hat{\bm k}),\\
{\bm d}_{\rm C}(\hat{\bm k})&=&\Delta_{{\rm A}_{1u}}\hat{\bm l}_{{\rm A}_{1u}}(\hat{\bm k}),
\end{eqnarray}
where $\Delta_{{\rm A}_{1u}}$ and $\Delta_{{\rm A}_{2u}}$ are the amplitudes for the $A_{1u}$ and $A_{2u}$ states, respectively.

The crucial difference between the $E_u$ state and the accidental scenario is that the latter retains the cubic symmetry.
This difference is manifested in the SNE in the A and C-phases.
Note that the order parameter of the $A_{1u}$ state is the same as that of the BW state, and thus realizes TRI TSCs in the C-phase.
As shown in the main text, the $A_{1u}$ state manifests finite SNC, $\alpha^{\sigma_z}_{xy},\; \alpha^{\sigma_x}_{yz}\; \alpha^{\sigma_y}_{zx}$.
In contrast with the anisotropic SNC in the topological superconducting state supported by $E_u$ state, the SNC in the $A_{1u}$ state satisfies $\alpha^{\sigma_z}_{xy}= \alpha^{\sigma_x}_{yz}= \alpha^{\sigma_y}_{zx}$.
The $A_{2u}$ state is a $f$-wave pairing and has point nodes along the (100), (010) and (001) directions.
These point nodes are Dirac points and related to helical $p$-wave pairing.
To see this more precisely, we focus on the point node in the (001) direction.
Near this point node, ${\bm k}\simeq (0,0,k_{\rm F})$, the order parameter for the $A_{2u}$ state becomes,
\begin{eqnarray}
\label{dvec2u}
{\bm d}_{\rm A}(\hat{\bm k})&=&\Delta_{{\rm A}_{2u}} \frac{\sqrt{35}}{2k_{\rm F}}(-k_x,k_y,0).
\end{eqnarray}
It is noted that Eq.~\eqref{dvec2u} is equivalent to the $d$-vector for the helical $p$-wave state described by ${\bm d}_{{\rm DSC},xy}$ with the opposite helicity.
Similarly, the other point nodes are also related to helical $p$-wave order parameters.
Specifically, the order parameter for the ${\rm A}_{2u}$ state is related to the helical state described with ${\bm d}_{{\rm DSC},yz}\;({\bm d}_{{\rm DSC},zx})$ in the vicinity of the point node in the $k_x\; (k_y)$-direction.
The discussions above successfully associate point nodes in $A_{2u}$ with the helical order parameters.
Hence, we conclude that all of $\alpha^{\sigma_z}_{xy},\; \alpha^{\sigma_x}_{yz}\; \alpha^{\sigma_y}_{zx}$ are finite in the $A_{2u}$ state.The finite spin-Nernst tensor elements are summarized in Table~\ref{ube13SNC}.

We conclude that the measurements of the spin-Nernst signal can demonstrate the helical superconducting order in U$_{1-x}$Th$_x$Be$_{13}$ and the anisotropy of the SNC in the different planes reflect the underlying pairing symmetry.

\begin{table}[hbtp]
  \caption{The finite spin-Nernst conductivity tensor elements in the A and C phases in each scenario.}
  \label{ube13SNC}
  \centering
  \begin{tabular}{lcr}
    \hline
      & the A phase & the C phase \\
    \hline \hline
    The degenerated scenario  & $\alpha_{xy}^{\sigma_z}$  & $\alpha_{xy}^{\sigma_z}\neq \alpha_{yz}^{\sigma_x}=\alpha_{zx}^{\sigma_y}$ \\
    The accidental scenario  & $\alpha_{xy}^{\sigma_z}=\alpha_{yz}^{\sigma_x}=\alpha_{zx}^{\sigma_y}$  & $\alpha_{xy}^{\sigma_z}=\alpha_{yz}^{\sigma_x}=\alpha_{zx}^{\sigma_y}$ \\
    \hline
  \end{tabular}
\end{table}

\bibliography{thermospin.bib}
\bibliographystyle{apsrev4-1_PRX_style}